\documentclass[twocolumn,english,aps,pre,showpacs,nofootinbib]{revtex4}
\usepackage[T1]{fontenc}
\usepackage[latin9]{inputenc}
\usepackage{amsmath}
\usepackage{amssymb}
\usepackage{esint}
\usepackage{graphicx}
\makeatletter
\@ifundefined{textcolor}{}
{%
 \definecolor{BLACK}{gray}{0}
 \definecolor{WHITE}{gray}{1}
 \definecolor{RED}{rgb}{1,0,0}
 \definecolor{GREEN}{rgb}{0,1,0}
 \definecolor{BLUE}{rgb}{0,0,1}
 \definecolor{CYAN}{cmyk}{1,0,0,0}
 \definecolor{MAGENTA}{cmyk}{0,1,0,0}
 \definecolor{YELLOW}{cmyk}{0,0,1,0}
 }

\usepackage{nicefrac}
\usepackage{dsfont}
\usepackage{xfrac}
\usepackage[normalem]{ulem}\@ifundefined{definecolor}
 {\usepackage{color}}{}

\makeatother

\usepackage{babel}
\begin{document}

\title{Conditions for positioning of nucleosomes on DNA}

\author{Michael Sheinman, Ho-Ryun Chung}
\address{Max Planck Institute for Molecular Genetics, 14195 Berlin, Germany}

\date{\today}
\begin{abstract}
Positioning of nucleosomes along eukaryotic genomes plays an important role in their organization and regulation. There are many different factors affecting the location of nucleosomes. Some can be viewed as preferential binding of a single nucleosome to different locations along the DNA and some as interactions between neighboring nucleosomes. In this study we analyzed how well nucleosomes are positioned along the DNA as a function of strength of the preferential binding, correlation length of the binding energy landscape, interactions between neighboring nucleosomes and others relevant system properties. We analyze different scenarios: designed energy landscapes and generically disordered ones and derive conditions for good positioning. Using analytic and numerical approaches we find that, even if the binding preferences are very weak, synergistic interplay between the interactions and the binding preferences is essential for a good positioning of nucleosomes, especially on correlated energy landscapes. Analyzing empirical energy landscape, we discuss relevance of our theoretical results to positioning of nucleosomes on DNA \emph{in vivo.}
\end{abstract}
\maketitle

\section{Introduction}
Our genome is packed and organized by nucleosomes---histone octamers wrapped around by $147$ bp of DNA \cite{kornberg1999twenty,richmond2003structure}. Nucleosomes are in some cases very well positioned while in others they are rather "smeared" along the DNA molecule \cite{gaffney2012controls,small2014single}. Their positioning properties are known to be important in the regulation of gene expression \cite{jiang2009nucleosome,bai2010gene,belch2010weakly}. There are many factors which determine positioning of nucleosomes along the DNA and their relative influence is a matter of active debate in the field (see Ref.~\cite{struhl2013determinants} for a review). 

The most discussed positioning mechanism is the DNA sequence heterogeneity. It is well known that to wrap DNA around a nucleosome one needs different energies for different DNA sequences \cite{trifonov1980sequence,satchwell1986sequence}. In this case the debate is only about the importance of DNA sequence preferences, relative to other factors. 

An obvious competitor of sequences preferences for nucleosomes positioning is thermal fluctuations. All measured binding energy differences between different sequences do not exceed a few $k_\text{B}T$ even for specially designed strongest binders, which do not exist in known genomes \cite{thaastrom1999sequence,thaastrom2004measurement,morozov2008extrinsic,takasuka2010direct,winkler2011histone}. This indicates that, at least in equilibrium, entropic forces are expected to play an important role. It is not entirely clear whether the nucleosomes reach (quasi)equilibrium and how they do it. It was suggested that some active chromatin remodeling enzymes, facilitate the equilibration of the nucleosomes by increasing the off-rate of the nucleosomes from the DNA \cite{padinhateeri2011nucleosome}. 
These and others active chromatin remodeling enzymes \cite{vignali2000atp,teif2009predicting} and DNA-binding proteins \cite{korber2004evidence} also affect nucleosomes positioning by actively moving the nucleosomes and by DNA binding competition.

In addition to external positioning signals, there are arguments for and evidences of interactions between neighboring nucleosomes along the DNA \cite{widom1992relationship,mergell2004nucleosome,muhlbacher2006tail,stehr2008effect,wang2008preferentially,grigoryev2009evidence,
liu2011influence,chereji2011statistical1,chereji2011statistical2,beshnova2014regulation}. The interactions are also expected to affect nucleosome positioning. This positioning factor is different from the one mentioned above, since it depends not on an absolute position of a nucleosome on the DNA, but on a relative position of two nucleosomes---the distance between two neighboring nucleosomes. 

In this study for simplicity we divide the positioning factors to two types. The first type includes all the factors which determine the position of a single nucleosome. One can characterize it by an effective binding energy landscape of a nucleosome along the DNA molecule that depends only on the location of the nucleosome along the DNA. The second type corresponds to interactions between neighboring nucleosomes. We characterize this positioning factor by an effective interaction potential that depends only on the distance between two neighboring nucleosomes. We assume that nucleosomes cannot invade each others DNA territories (although it is not entirely true \cite{engeholm2009nucleosomes,mobius2013toward,chereji2014ubiquitous} this is not expected to affect significantly the conclusions of this study). In addition we analyze only the equilibrium distribution of nucleosomes, ignoring non-equilibrium aspects.
 
In general, we address the following question: within the framework of the above assumptions, what should be the properties of the effective energy landscape and effective interaction potential between neighboring nucleosomes to achieve good positioning? 
We analyze energy landscapes with different properties and different interaction potentials and derive conditions leading to good positioning of nucleosomes. We show that the interactions between nucleosomes can significantly improve their positioning even on almost flat and highly correlated energy landscapes. In this case, if the positioning is good, one expects to observe also large length-scale fluctuations of nucleosome occupancy along the DNA. Comparing our results to empirical study, we find good qualitative and quantitative agreement. 

\begin{figure*}[!]
\centering
  \includegraphics[width=\textwidth]{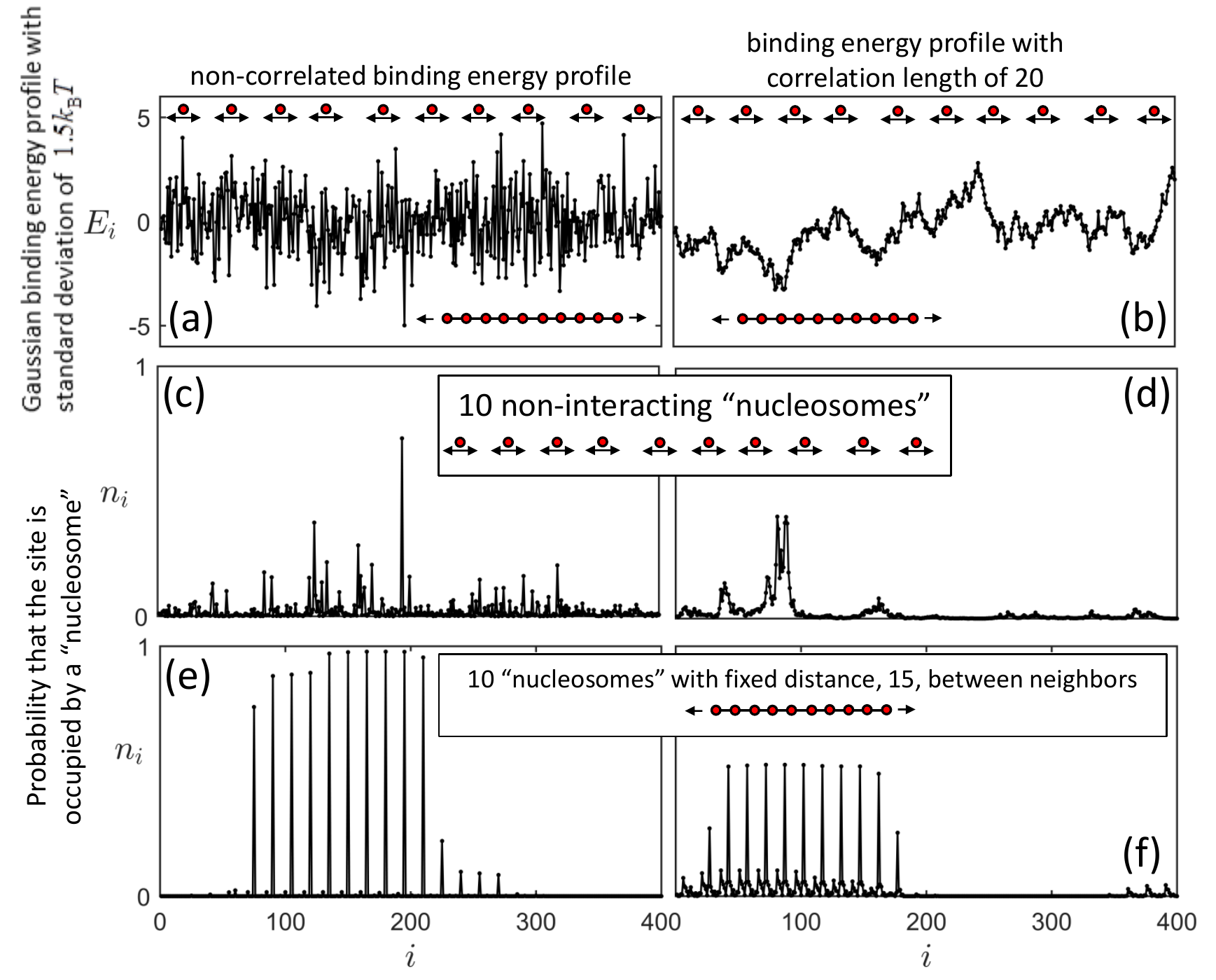}
\caption{Illustration of how interactions between nucleosomes improves their positioning along the DNA. In this toy example "nucleosomes" are merely particles of 1bp size with disordered binding energy landscape on 400bp-long, circular "DNA". (a) Binding energies of a single nucleosome are i.i.d random variables with a Gaussian distribution with the standard deviation of $\sigma=1.5k_\text{B}T$. (b) Binding energies of a single nucleosome  are normally distributed with the standard deviation of $\sigma$ and are correlated, such that 	$\left<E_i E_{i+r}\right>=\sigma^2e^{-\frac{r}{r_c}}$, with $r_c=20$. (c) The probability that the site $i$ is occupied by a nucleosomes, $n_i$ is plotted vs. $i$ for $10$ non-interacting nucleosomes located on the binding energy profile from (a). (d) The same as (c) but for the binding energy profile from (b). (e) The same as (c) but for nucleosomes with strong interactions, such that the distance between two neighboring nucleosomes is constrained to 15bp. (f) The same as (e) but for the binding energy profile from (b).}
\label{Illustration}
\end{figure*}

Before we start with detailed description of the model and its analysis, it might be instructive to illustrate the main message of the paper with a toy example. Consider non-correlated Gaussian binding energy landscape with standard deviation of $1.5k_{\text{B}}T$, on a circular DNA of length $400$bp (see Fig. \ref{Illustration}(a)). Ten non-interacting "nucleosomes" of size $1$bp cannot be well positioned with such a weak energetic disorder (see Fig. \ref{Illustration}(c)). However, adding very strong interaction between the nucleosomes, such that the distance between them is restricted to $15$bp, one get good positioning on the same, weak energetic profile (see Fig. \ref{Illustration}(e)). Autocorrelation of an energy (see Fig. \ref{Illustration}(b)) makes the positioning even more problematic (see Fig. \ref{Illustration}(b)). However, again, strong interactions between the nucleosomes improves it to a reasonable level (see Fig. \ref{Illustration}(d)). In the paper we will derive, within quite general set of assumptions, conditions for positioning on uncorrelated and correlated correlated binding energy landscapes. In Figs. \ref{Illustration}(e) and (f) one can see that, when interactions between the nucleosomes are exploited for a better positioning, there are long length-scale fluctuations of the nucleosomes occupancy along the DNA---there are long enriched and long depleted regions. Below we analyze the properties of such regions and demonstrate relevance of this effect to empirical data.

The structure of the paper is as following. In Section \ref{The model} we formulate the model we use. In Section \ref{Quantities of interest} we define the quantities we use to characterize positioning of nucleosomes on the DNA. In Section \ref{Positioning of one nucleosome} we analyze positioning of a single nucleosome. The purpose of this Section is not only didactic, because we use its results below.
In Section \ref{Positioning of multiple nucleosomes with only hardcore interactions} we analyze positioning of a many nucleosomes with only hard-core interactions. The purpose of this Section is to contrast it to the case with interactions between neighboring nucleosomes and demonstrate the importance of these interaction in Section \ref{Positioning of strongly interacting nucleosomes}. We generalize our results for energy landscapes with autocorrelation in Section \ref{Energy landscape with correlations}. In Section \ref{Relevance to empirical results} we discuss relevance of our conclusions to real systems and compare to empirical results. After discussion about tunability and robustness of positioning issues, emerging from our results, in Section \ref{Tunability and robustness of the positioning}, we summarize in Section \ref{Summary}.
We proceed now with a detailed description of our model.

\section{The model}
\label{The model}
We analyze the following lattice based model, which is often used to calculate occupancy of DNA-binding proteins \cite{teif2010statistical}. In a grand-canonical ensemble on average $N$ nucleosomes are located on a linear DNA of length $L$, in units of bp with reflecting boundaries. Note, that near the saturation of the DNA by nucleosomes the grand-canonical and canonical ensembles may be different \cite{woodbury1981free,teif2009predicting}.
Each nucleosome occupies $W=147$bp on the DNA, such that if its leftmost position is bound to a site $i$ another nucleosome cannot bind with its leftmost position to any of the sites in the interval $[i-W+1,i+W-1]$. Due to DNA sequence preferences of nucleosomes or any other reason a nucleosome bound with its leftmost position to site $i$ possesses an energy $E_i$. In addition to this energy there is an interaction energy between two neighboring nucleosomes. Given the distance $r \geq 0$ between the leftmost positions of the two nucleosomes the interaction energy is given by $V(r)$. The hardcore interaction is realized by $V(r)=\infty$ for $0 \leq r < W$. To obtain the equilibrium properties of the nucleosomes we numerically solve the recursive equation for the partition function \cite{gurskii1977precise,teif2012calculating}.

Our focus is the following question: what should be the properties of the signal in the one-nucleosome energy profile along the DNA, $E_i$ and the interaction between the nucleosomes, $V(r)$, to achieve good positioning of nucleosomes on the DNA? In the next Section we define this question in more quantitative terms.


\section{Quantities of interest}
\label{Quantities of interest}
In this paper we focus on several quantities which reflect positioning of nucleosomes. Each one of them can be derived from an equilibrium probability of the site $i$ to be covered by the leftmost position of a nucleosome, $n_i$ (start site probabilities). The average number of nucleosomes, $N$, is given by
\begin{equation}
	N=\sum_{i=1}^L n_i.
\end{equation}

We also define an ordered vector of occupancies, $n^\text{o}_m$, such that $n^\text{o}_1$ is the occupancy of the most occupied site (site with the highest value of $n_i$), $n^\text{o}_2$ is the occupancy of the second-most occupied site etc.

For cases when it is \emph{not} important how a base-pair along the DNA is covered by a nucleosome (by which part of the nucleosome it is covered) the occupancy function
\begin{equation}
	\rho_i=\sum_{j=0}^{W-1} n_{i-j}
\end{equation}
is of an interest. We define the average coverage of the DNA by
\begin{equation}
	\rho=\frac{1}{L}\sum_{i=0}^{L} \rho_i=\frac{NW}{L}.
\end{equation}

There are different ways to define a measure of how well nucleosomes are positioned along the DNA. In this paper we use a very simple one: given that there are $N$ nucleosomes, we define as $\mathcal{P}$ the fraction of nucleosomes which occupies $N$ most occupied locations along the DNA. Namely,
\begin{equation}
	\mathcal{P}=\frac{\sum\limits_{m=1}^N n^\text{o}_m}{\sum\limits_{m=1}^L n^\text{o}_m}=\frac{1}{N} \sum_{m=1}^N n^\text{o}_m.
	\label{MathCalP}
\end{equation}  
In the case of a single nucleosome, $N=1$ this definition becomes simply the occupation probability of the ground state---the order parameter of, say, the Random Energy model \cite{derrida1980random}. For multiple nucleosomes $\mathcal{P}$ can be viewed as a fraction of well positioned nucleosomes. In the case of a perfect positioning $\mathcal{P}=1$, while in case of positioning whatsoever, $\mathcal{P}=N/L \ll 1$. The last quantity, $N/L$, can be at most $1/W$ for $\rho=1$. 

As we show below, for correlated energy landscapes, in some cases nucleosomes are not positioned well on a single bp length scale but are positioned well within a few basepairs. In this case the value of $\mathcal{P}$ does not characterize fully the positioning goodness. For those case we exploit the following generalization of $\mathcal{P}$. Denoting the profile around the $m$'th largest values of $n_i$, as $n^\text{o}_m(s)$
we define for odd values of $k$
\begin{equation}
	\mathcal{P}_k=\frac{1}{N} \sum_{m=1}^N  \sum_{s=-(k-1)/2}^{(k-1)/2} n^\text{o}_m(k).
	\label{Pr}
\end{equation}
The value of $\mathcal{P}_k$ is the measure of positioning given that one does not care about fuzziness on the lengthscale of $k$. One can easily see that on the level of one bp resolution $\mathcal{P}_1$ is given by $\mathcal{P}$. However, as we show below, on correlated energy landscapes and/or with interaction potential with wide wells the function $\mathcal{P}_k$ can be much more informative than its single-bp resolution value $\mathcal{P}_1=\mathcal{P}$.
We turn now to the consideration of positioning for different scenarios, starting from the simplest one.

\section{Positioning of one nucleosome}
\label{Positioning of one nucleosome}
It is instructive to consider first the simplest case of a single nucleosome, $N=1$, on a DNA of a certain length, $L$. To position it on the DNA in equilibrium one should generate non-uniform energy profile along the DNA. Below we discuss possible energy profiles which can be roughly divided to designed and generic ones. 
 
\subsection{Designed energy landscape}
Conceptually, the easiest way to position a nucleosome on a site $j$ is to design an energy landscape such that up to an additive constant $E_{i=j}=-E$ and $E_{i \neq j}=0$. In this designed DNA case the positioning measure \eqref{MathCalP} is given by (we measure all energies in units of $k_{\rm B} T$)
\begin{equation}
	\mathcal{P}=n_1^\text{o}=\frac{1}{1+Le^{-E}},
\end{equation}
such that the nucleosome occupies site $i$ with probability of order one if $E \gtrsim \ln L$. Therefore, having in the arsenal only energies of the order of a few $k_B T$ one can position a single nucleosome only on short sequences of tens---hundreds base-pairs, even on the best possible energy landscape.

\subsection{Disordered energy landscape with Gaussian distribution}
In the more generic case of a disordered energy landscape the problem can be mapped to the Random Energy model \cite{derrida1980random,gerland2002physical,slutsky2004kinetics,sheinman2012classes}. In this case the probability that the leftmost location of the nucleosome is located on site $i$ along the DNA is given by
\begin{equation}
n_i=\frac{e^{-E_i}}{Z},
\end{equation} 
where the partition function is
\begin{equation}
Z=\sum_{i=1}^{L}e^{-E_i}
\end{equation} 
Therefore, to position a nucleosome on site $i$ one has to fulfill the condition $e^{-E_i} \sim Z$. Consider the case where the energies $\{ E_i\}$ are a set of i.i.d random variables with a normal probability distribution with standard deviation $\sigma$:
\begin{equation}
	\Pr\left( E_i \right)=\frac{e^{-\frac{E_i^2}{2\sigma^2}}}{\sqrt{2\pi \sigma^2}}.
	\label{Gauss}
\end{equation}
In this case the lowest energy, $E_1^\text{o}$, is well approximated by 
\begin{equation}
\int_{-\infty}^{E_1^\text{o}}  \Pr\left( E \right) dE = \frac{1}{L}
\label{E1eq}
\end{equation}
The solution is given by
\begin{equation}
E_1^\text{o} \simeq  \sigma  \sqrt{2}   {\rm erf}^{-1}\left(\frac{2 }{L}-1\right)  \simeq -\sigma \sqrt{2 \ln L}.
\label{E1}
\end{equation}
In the limit of zero temperature (or, equivalently infinite disorder strength $\sigma$) the nucleosome will occupy the state with the lowest energy (ground state) with probability 1. However, for small non-zero temperatures non-ground states will be partly occupied, such that the occupation probability of the ground state is smaller than one. Consider the $m$-lowest energy ($1$-lowest energy means the lowest one, $2$-lowest energy is the second lowest energy, etc.) on the DNA,  $E_{m}^\text{o}$. Its value can be well approximated by
\begin{equation}
\int_{-\infty}^{E_m^\text{o}}  \Pr\left( E \right) dE = \frac{m}{L}
\end{equation}
The solution is given by
\begin{equation}
E_m^\text{o} \simeq   \sigma  \sqrt{2}   {\rm erf}^{-1}\left(\frac{2 m}{L}-1\right) \simeq E_1^\text{o}+\frac{\sigma}{\sigma_{\rm f}}\ln m.
\end{equation}
where the freezing disorder strength is given by
\begin{equation}
\sigma_{\rm f} \simeq - \sqrt{2}   {\rm erf}^{-1}\left(\frac{2 m}{L}-1\right) \simeq \sqrt{2 \ln L}.
\end{equation}
The affinities of the $m$-lowest state are given by
\begin{equation}
K^\text{o}_m=e^{-E^\text{o}_m} \simeq e^{-\sqrt{2}   {\rm erf}^{-1}\left(\frac{2 m}{L}-1\right)} \simeq \frac{K_1^\text{o}}{m^{\sigma/\sigma_{\rm f}}}.
\end{equation}
In the low temperature (high disorder) limit, $\sigma>\sigma_{\rm f}$ the occupation of the lowest state is given by
\begin{equation}
\mathcal{P}=n_1^\text{o} \simeq \frac{K^\text{o}_1}{\sum_m K^\text{o}_m} = \frac{1}{\zeta\left( \frac{\sigma}{\sigma_{\rm f}} \right)} \simeq \left(1+\frac{2^{1-\frac{\sigma}{\sigma_{\rm f}}}}{\frac{\sigma}{\sigma_{\rm f}}-1}\right)^{-1},
\label{Pbelow}
\end{equation}
where $\zeta(s)=\sum_{m=1}^\infty m^{-s}$ is the Riemann zeta function. Above the freezing point $\sigma<\sigma_{\rm f}$ the sum in the equation above diverges. In this case the annealed approximation of the free energy is valid and the partition function is not widely distributed around its mean value
\begin{equation}
	Z\simeq\left< Z \right>=Le^{\sigma^2/2}.
\end{equation}
In this regime the probability of occupation of any site is given by
\begin{equation}
\mathcal{P}=n_1^\text{o}  = \frac{e^{-E_m^\text{o}}}{Z} \simeq \frac{e^{\sigma\sqrt{2 \ln L}}}{Le^{\sigma^2/2}},
\label{Pabove}
\end{equation}
such that it vanishes in the thermodynamic limit, $L \rightarrow \infty$.
In genomes of lengths in the range $
L=10^6-10^9$ bp the freezing transition happens in the range $\sigma_{\rm f} \simeq \sqrt{2 \ln L}=5.3-6.4 k_{\rm B}T$.

In sum positioning of a single nucleosome is determined by the disorder strength. It is well positioned on DNA of length $L$ (such that $\mathcal{P} \simeq 1$) with energetic Gaussian, uncorrelated disorder with width $\sigma$ for $\sigma \gg \sqrt{2 \ln L}$  and is "smeared" along the DNA (such that $\mathcal{P} \ll 1$) in the opposite limit of weak disorder, $\sigma \ll \sqrt{2 \ln L}$. In Fig.~\ref{OneNucleosomeP} the above considerations are illustrated. In Appendix \ref{Appendix Positioning of one nucleosome} we discuss positioning on energy landscape with non-Gaussian distributions. Now we turn to discuss positioning of many nucleosomes on the DNA.

\begin{figure}[h!]
\centering
  \includegraphics[width=0.5\textwidth]{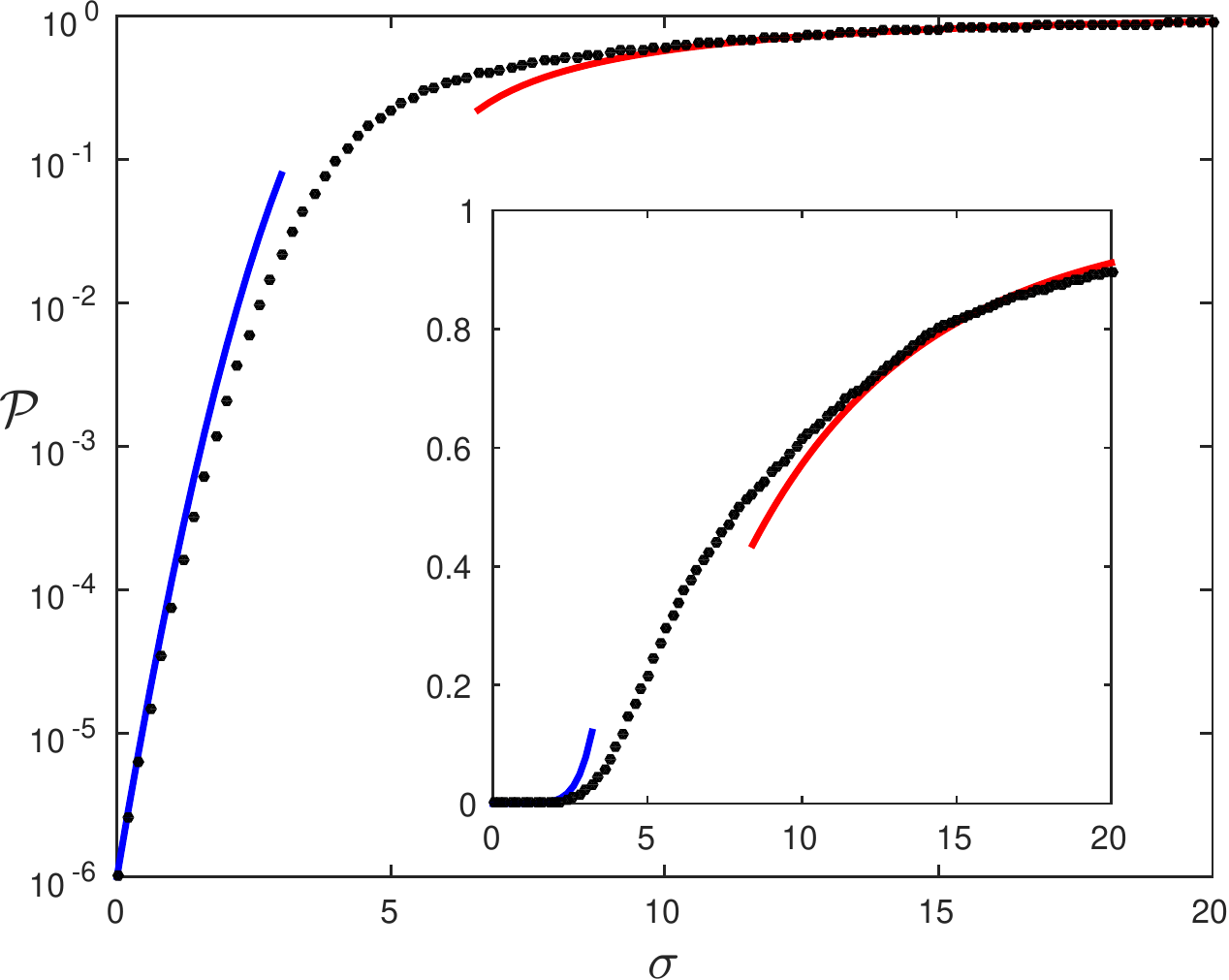}
\caption{Positioning of a single nucleosome on the DNA with length $L=10^6bp$, such that the freezing transition is at $\sigma_{\rm f} \simeq 5.3k_{\text{B}}T$. Occupancy of the deepest energy well is plotted vs. disorder width. The dots represent numerical simulation---median of $100$ realizations of the disordered energy landscape energy. The lines represent the analytic solution: Eq.~\eqref{Pbelow} for $\sigma > \sigma_{\rm f}$ and Eq.~\eqref{Pabove} for $\sigma < \sigma_{\rm f}$. Inset: the same plot in the linear scale.}
\label{OneNucleosomeP}
\end{figure}

\section{Positioning of multiple nucleosomes with only hardcore interactions}
\label{Positioning of multiple nucleosomes with only hardcore interactions}
Here we analyze positioning of $N \gg 1$ nucleosomes with only hard-core interactions. The study of particles with hard-core repulsions has a long history and is relevant to many applications. In the context of protein-DNA binding such a repulsion between proteins leads to crowding and influences the binding properties \cite{mcghee1974theoretical,li2009effects,morelli2011effects,sheinman2012does}. Here we consider positioning of nucleosomes of a finite size, $W \geq 1$ \cite{epstein1978cooperative,kornberg1988statistical} and focus on positioning properties on different energy landscapes.
\subsection{Designed energy landscape}
\begin{figure}[h!]
\centering
  \includegraphics[width=0.5\textwidth]{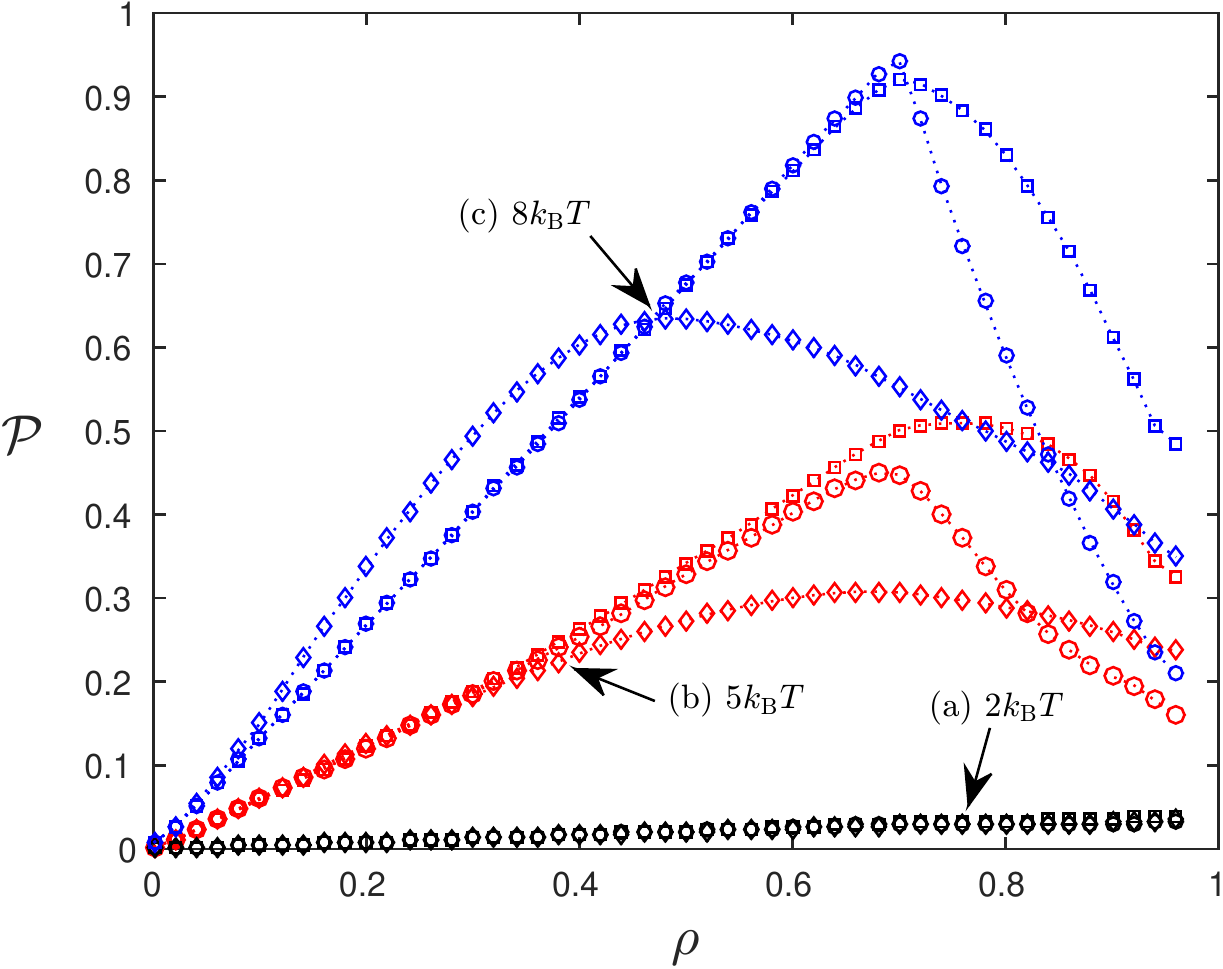}
\caption{Numerical results for positioning of nucleosomes with only hard-core interactions with $W=147bp$ on designed energy landscapes of a length of $L=10^3 \times W$. Positioning parameter is plotted as a function of the coverage fraction $\rho=NW/L$. The energy profile is designed such that neighboring energy wells are separated by a distances with the following properties: Each distance is $W+63bp$ (circles), any distance is a sum of $W$ and a number drawn from a geometric distribution with an average of $63bp$ (squares) and each distance is a number drawn from a geometric distribution with an average of $W+63bp$ (diamonds). The depths of the energy wells relative to the rest positions on the DNA are: (a) $2$, (b) $5$ and (c) $8$ $k_{\rm B}T$. The lines are to guide the eye.}
\label{HardCoreDesignP}
\end{figure}

Consider first a designed case when there are $N$ nucleosomes of size $W$ on a DNA of length $L$ and $N$ energy wells of energy $-E$ while the rest of the DNA positions have zero energy. For the best positioning all the distances have to be larger or equal to $W$, such that nucleosomes don't have to overlap to occupy all the energy wells. In this case to position well the nucleosomes one needs $E \gtrsim \ln \frac{L}{N}$. Then, having, say, $10-70-90\%$ coverage of the DNA by nucleosomes of length $W=147$, to position the nucleosomes one needs energy well to be deeper than $7-4-3k_BT$. Moreover, even if the wells are that deep but the number of nucleosomes differs from the number of energy wells the positioning is getting worse. 
To make the positioning more robust one can make the distance between the wells being random. However, doing this one has to keep the minimal distance between two neighboring wells to be $W$. Otherwise, the nucleosomes, being not able to overlap, spread more on the DNA decreasing the positioning parameter, $\mathcal{P}$ (see Fig.~\ref{HardCoreDesignP}). In a more generic, disordered case the positioning is more problematic. We turn now to discuss this case.

\subsection{Disordered energy landscape with Gaussian distribution}
\label{No Interaction Generically disordered energy landscape}
Consider positioning of $N$ nucleosomes on uncorrelated disordered energy profile normally distributed with standard deviation $\sigma$ (see Eq.~\eqref{Gauss}). In the regime $\sigma \ll \sqrt{2\ln \frac L N }$ the nucleosomes are poorly positioned, while in the opposite regime,
\begin{equation}
	\sigma \gg \sqrt{2\ln \frac L N },
	\label{NoIntPos}
\end{equation}
the positioning is good. In sum, having, say, $10-70-90\%$ coverage of the DNA by nucleosomes of length $W=147$, to position the nucleosomes one needs disorder strength, $\sigma$ to be larger than $3.8-2.9-2.4k_BT$.

\begin{figure}[h!]
\centering
  \includegraphics[width=0.5\textwidth]{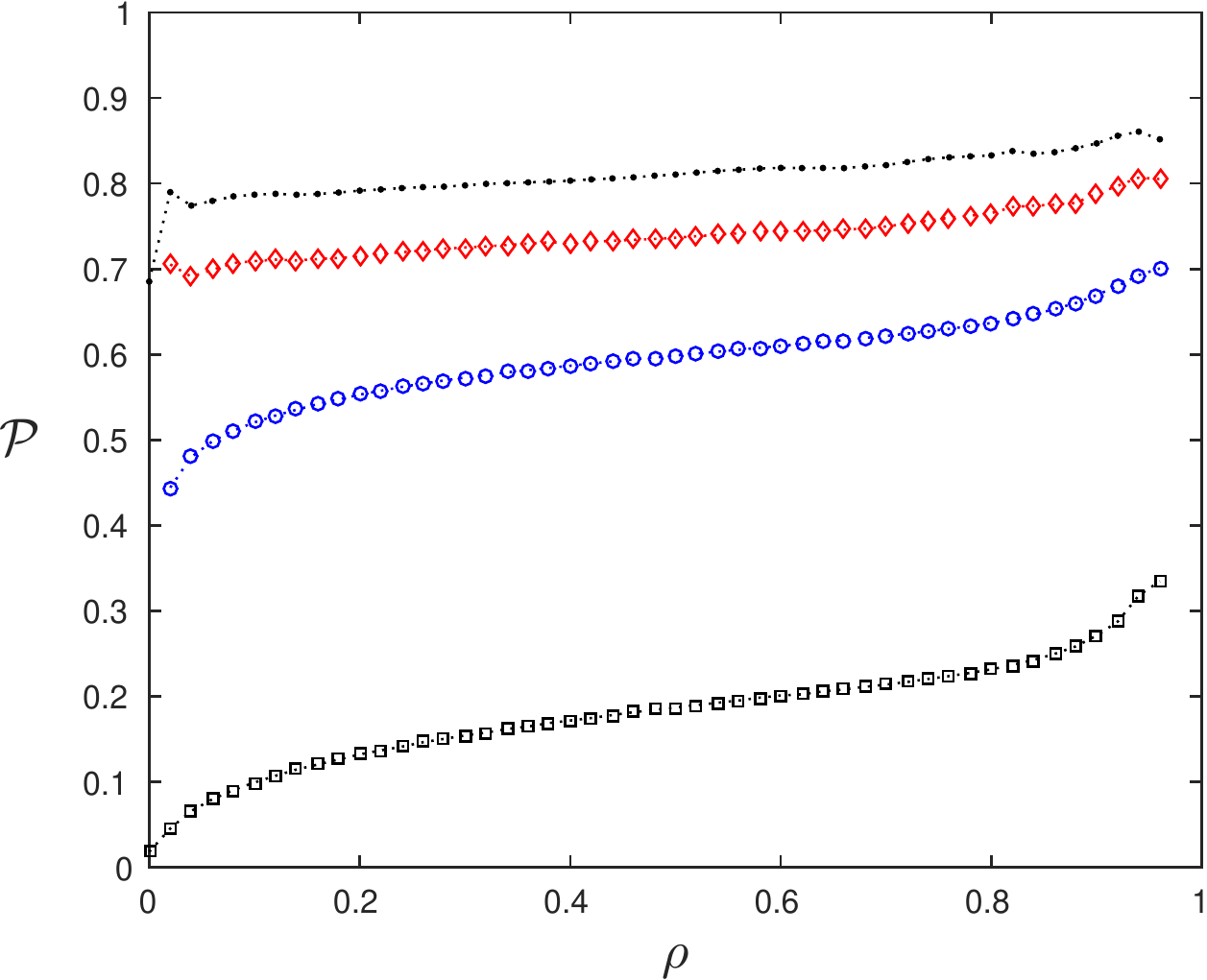}
\caption{Positioning of nucleosomes with only hard-core interactions. Entropy of the DNA with length $L=10^3 \times W$ and $W=147$ as a function of the coverage fraction $\rho=NW/L$. The markers represent numerical simulation for $\sigma=2$ (squares), $5$ (circles), $8$ (diamonds) and $11$ (dots) $k_{\text{B}}T$.}
\label{HardCoreGenericP}
\end{figure}

The derived requirement for positioning may sound weak. However, in fact it means that, say, for $\rho=70\%$ and $\sigma=5 k_\text{B}T$ (moderate positioning regime, $\mathcal{P} \simeq 0.6$, as shown in  Fig.~\ref{HardCoreGenericP}) the typical energy well for a nucleosome is $14 \pm 2 k_\text{B}T$ deep (see Eq.~\eqref{E1} with $L$ replaced by $L/N$), relative to a random DNA sequence. 
In sum, one can see that without interactions the energy variations required for a good positioning seem to be above the ones measured in experiments \cite{thaastrom1999sequence,thaastrom2004measurement,morozov2008extrinsic,takasuka2010direct,winkler2011histone}. 
In Appendix \ref{Appendix Positioning of multiple nucleosomes with only hardcore interactions} we discuss positioning of nucleosomes with only hard-core interaction on energy landscape with non-Gaussian distributions and show that the results in this Section do not change qualitatively in this case. 
In the next section we show how interactions between nucleosomes allow to position them with much weaker energetic disorder along the DNA.

\section{Positioning of strongly interacting nucleosomes}
\label{Positioning of strongly interacting nucleosomes}
Consider $N$ nucleosomes on a DNA of length $L$. As we show above, weak sequence specificity cannot position nucleosomes. In this Section we analyze positioning of interaction nucleosomes on designed and disordered energy landscapes. 

Interaction between neighboring nucleosomes, were suggested before as one of the driving forces, ordering nucleosomes \cite{lubliner2009modeling,chereji2011statistical1,chereji2011statistical2,beshnova2014regulation}.
Here, we consider for simplicity the minimal model of an interaction with an energy well when two neighboring nucleosomes are at a distance $R - \Delta \leq r \leq R + \Delta$ from each other. Namely, the interaction potential is of the form
\begin{equation}
	e^{-V(r)}= 
	\begin{cases} 
	0 &  r<W  \\
	e^v=\kappa & R+\Delta\geq r \geq R-\Delta \geq W \\
	1 &  \rm{else}. 
	\end{cases}
	\label{4}
\end{equation}
For simplicity we analyze narrow interaction energy well of only one bp, $\Delta=0$. Further we discuss other possible potential functions in general and, in particular, importance of a finite width of the interaction potential well, $\Delta>0$.

\begin{figure}[h]
\centering
  \includegraphics[width=0.5\textwidth]{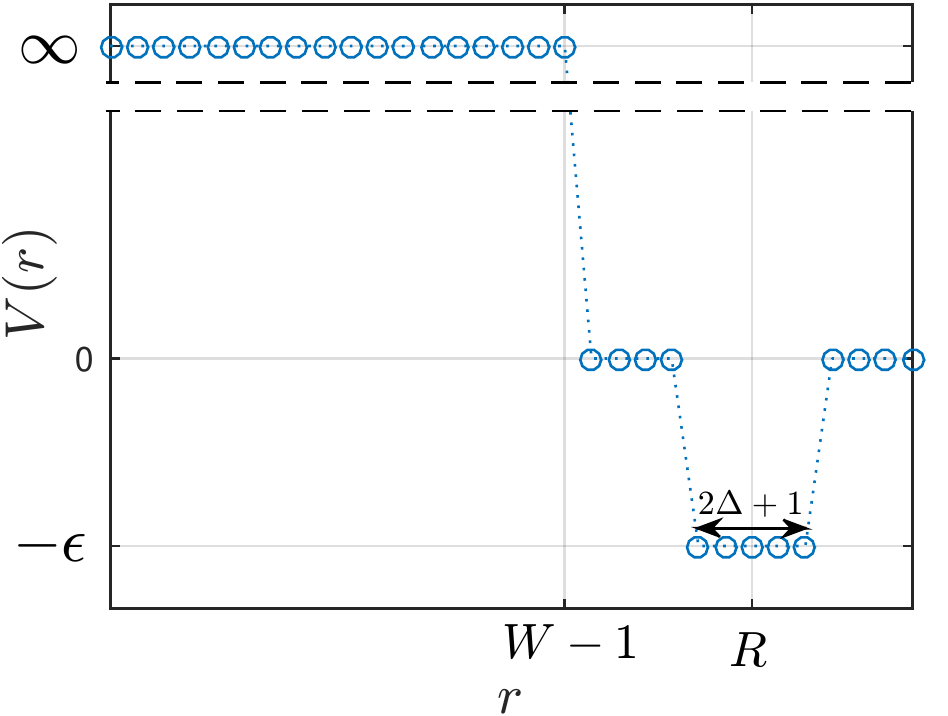}
\caption{Illustration of the interaction strength between neighboring nucleosomes used in the paper. $V(r)$ from Eq.~\eqref{4} is presented. In this particular case $R=154$, $W=147$ and $\Delta=2$.} 
\label{Potential}
\end{figure}

For strong enough interactions strength (large values of $\kappa$) the nucleosomes gather to clusters, such that in each cluster the distance between the neighboring nucleosomes is $R$. As we show below this clustering effect can significantly improve positioning of nucleosomes. We demonstrate it first on a designed energy landscape.

\subsection{Designed energy landscape}

In order to exploit interactions between nucleosomes and position them on a weak but designed energy landscape one should have a spatial resonance between the energy wells distance and the preferable distance between neighboring nucleosomes, $R$. Consider first the case of periodic array of wells with energy $-E$, such that the affinity is $K=e^E$, and set the number of nucleosomes to be the number of wells. Due to interactions nucleosomes locally crystallize to ordered arrays with nearest-neighbors distance of $R$. Denoting the average length of a cluster by $M$, one gets the condition for the positioning of the cluster (and, therefore, all the nucleosomes in the cluster):
\begin{equation}
	ME \gg \ln \frac{L}{N/M}
\end{equation}
or
\begin{equation}
	\frac{K^M}{M} \gg \frac{L}{N}.
	\label{PosDesign}
\end{equation}
The average number of nucleosomes in a cluster, $M$, is given by
\begin{equation}
	M=\frac{1}{1-\Pr(r=R)},
	\label{MP}
\end{equation}
where $\Pr(r=R)$ is the probability that the distance between two neighboring nucleosomes is $R$. These quantities can be calculated using the self-consistency equation:
\begin{equation}
	\Pr(r=R)=1-\frac{1}{M}=\frac{K \kappa^2}{{K\kappa^2}+\frac{L}{N/M}}.
\end{equation} 
The solution is given by
\begin{equation}
	M=
		\begin{cases} 
		\kappa\sqrt{K\frac{N}{L}} &  \kappa\sqrt{K\frac{N}{L}} \gg 1  \\
		1 & \kappa\sqrt{K\frac{N}{L}} \ll 1 
	\end{cases}.
	\label{M}
\end{equation}

Therefore, satisfying condition \eqref{PosDesign}, one gets strong improvement of the positioning ($\mathcal{P}\simeq1$) even with very weak wells $K \simeq 1$. However, if the wells do not have a good periodicity or their period is different from the preferential distance between the nucleosomes the interactions do not improve the positioning.

In sum, one can position strongly interacting nucleosomes on a designed energy landscape with shallow energy wells. However, this positioning is not robust to change of nucleosome coverage fraction and requires fine tuning of the distances between energy wells along the DNA. We turn now to discuss positioning on generic energy landscapes where the positioning is not so strong but is more robust to properties of the energy landscape and the interaction potential.


\subsection{Disordered energy landscape with Gaussian distribution}
In the case when the energy along the DNA is a set of independent normally distributed variables with standard deviation $\sigma$  (see Eq.~\eqref{Gauss}), local crystallization of interacting nucleosomes also plays a major role in positioning for small values of $\sigma$. As shown in Section \ref{No Interaction Generically disordered energy landscape}, with only hard-core interactions the positioning is possible only when condition \eqref{NoIntPos} holds. Here we discuss the opposite limit of weak disorder and show that interactions between the nucleosomes can position nucleosomes even in this case. Consider a cluster of crystallized nucleosomes. The effective energy landscape for such a cluster possesses a stronger disorder than for an individual nucleosome. Namely, for a cluster of, say, $m$ nucleosomes the standard deviation of cluster's total energy distribution is $\sqrt{m} \sigma$, where $\sigma$, as before, is the energy standard deviation of a single nucleosome. However, for $m \gg 1$, this effective energetic disorder possesses an approximate periodicity of $R$ because shifting the cluster by this length the total energy of the cluster does not change much. Nevertheless, local  crystallization, increasing the effective disorder strength relative to the one for a single nucleosome, causes the positioning of clusters and, therefore, positioning of individual nucleosomes.

Consider a single typical cluster of a size $M \gg 1$. Typical available space for it is given by $\frac{L}{N/M}$, while the typical minimal energy is given by $\sqrt{M}\sigma \sqrt{2\ln R}$. Therefore, it will be frozen if  
\begin{equation}
	\sqrt{M} \sigma \gg \sqrt{2 \ln R}
	\label{srCond}
\end{equation}
and will be "smeared" on its available space in the opposite limit.
 
\begin{figure}[h!]
\centering
  \includegraphics[width=0.5\textwidth]{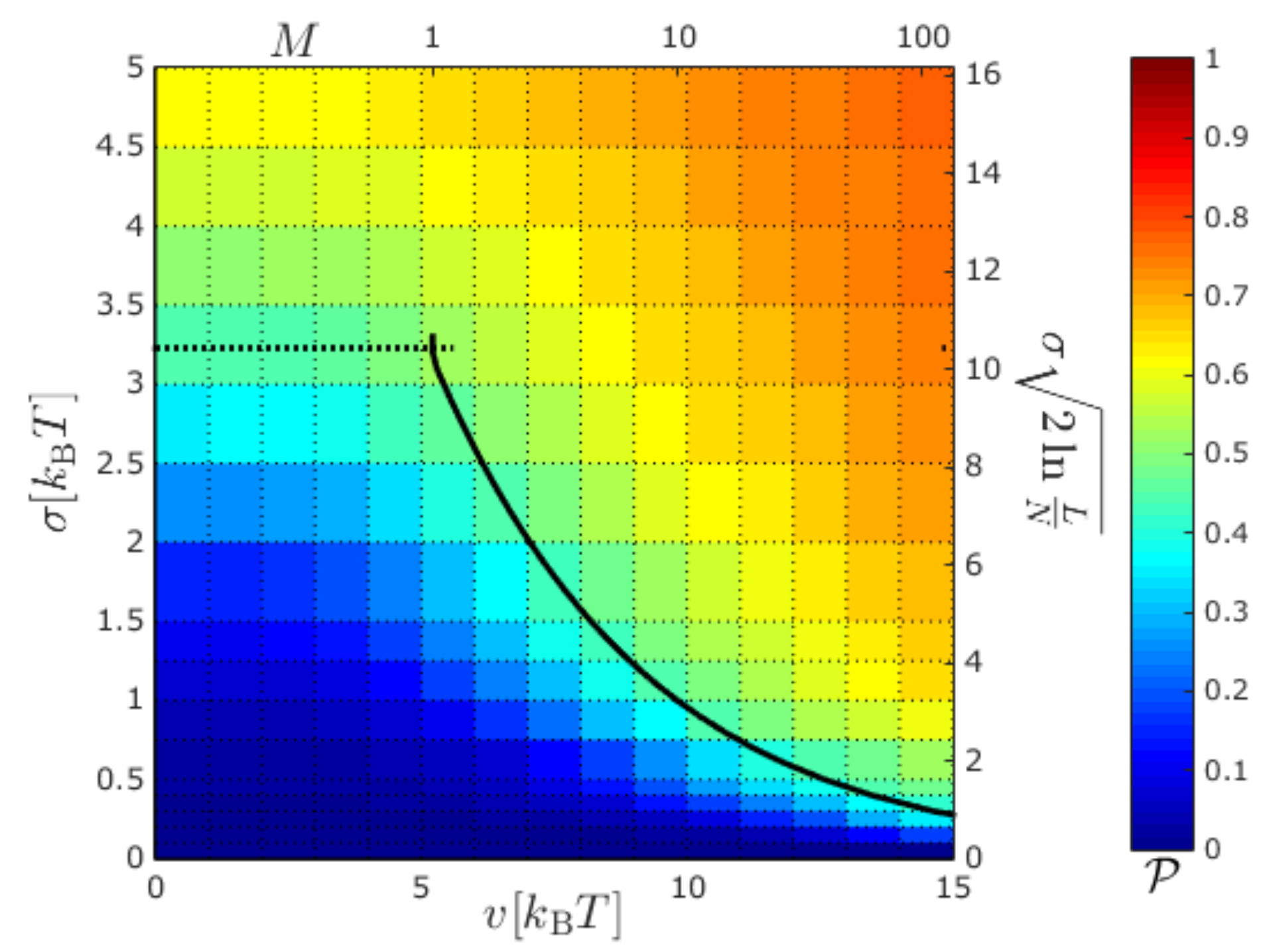}
\caption{Positioning of nucleosomes with interactions on Gaussian energy landscape. Positioning parameter, $\mathcal{P}$ for the DNA with length $L=10^4 \times W$ and $W=147$ for the coverage fraction $\rho=NW/L=80\%$ is plotted vs. disorder strength (left axis) and interaction strength (bottom axis) with preferable distance of $R=148$. On the top one can see the average size of the crystallized cluster of nucleosomes, derived from Eq.~\eqref{M}. On the right the typical binding energy of a nucleosome (relative to the average energy) is shown.
 The lines represent the analytic conditions for a good positioning, Eqs.~\eqref{NoIntPos} (dotted line) and \eqref{GenereicIntCond} (solid line).} 
\label{PhaseDiagramNormal}
\end{figure}

The value of $M$ can be estimated in the following way, using Eq.~\eqref{MP}:
\begin{equation}
	\Pr(r=R)=1-\frac{1}{M}=\frac{\kappa}{{\kappa}+\frac{L}{N/M}}.
\end{equation} 
Thus, 
\begin{equation}
	M=
		\begin{cases} 
		 \sqrt{\kappa\frac{N}{L}} &  \kappa\gg \frac{L}{N}   \\
		1 & \kappa\ll \frac{L}{N} 
	\end{cases}.
	\label{M}
\end{equation}
Combining Eqs.~\eqref{srCond} and \eqref{M}, the required strength of interactions to position nucleosomes is given by
\begin{equation}
	\kappa \gg \frac{4 }{\sigma^4 }  \frac{W}{\rho}\ln^2 R.
	\label{GenereicIntCond}
\end{equation}
In the case when $\sigma \gg \sqrt{2\ln R }$ condition \eqref{GenereicIntCond} has to be replaced by $\kappa\gg \frac{L}{N}$. However, in this case condition \eqref{NoIntPos} is satisfied, such that the positioning is possible with only hard-core interactions. Thus, as shown in Fig.~\ref{PhaseDiagramNormal}, if at least one of the conditions \eqref{NoIntPos} and \eqref{GenereicIntCond} holds the positioning is good, such that $\mathcal{P} \simeq 1$, while otherwise $\mathcal{P} \ll 1$.
One can see that strong interactions between nucleosomes are able to improve their positioning. In Appendix \ref{Appendix Positioning of strongly interacting nucleosomes} we discuss positioning of strongly interacting nucleosomes on energy landscape with non-Gaussian distributions and show that the results in this Section do not change qualitatively in this case.

The described local clustering of nucleosomes not only improves positioning of nucleosomes but also has another consequences---large length scale fluctuations of occupancy. We turn now to discuss this aspect of interactions-assisted positioning of nucleosomes.

\subsection{Large-scale fluctuations of occupancy}
\label{Large-scale fluctuations of occupancy}
Apart from the positioning of nucleosomes on small length-scale, there is another feature that is highly influenced by interactions between neighboring nucleosomes--- the large-scale fluctuations of occupancy of nucleosomes. One can also interpret this effect as long length scale positioning.
Without interactions occupancy $\rho_i$ averaged over thousands of base-pairs is not expected to deviate significantly from its average value, $\rho$. However, if neighboring nucleosomes possess a preferential distance, $R$ which is smaller than the average linker length, $W/\rho$, and the nucleosomes are very well positioned one gets long DNA regions which are enriched by nucleosomes and, therefore, regions depleted with nucleosomes. The correlation length of the occupancy on enriched regions for strongly interacting and very well positioned nucleosomes is given (using Eq.~\eqref{M}) by
\begin{equation}
	RM \sim R\sqrt{\kappa\frac{\rho}{W}}. 
\end{equation}
On this length scale the nucleosomes are enriched and their mean occupancy is given by $W/R$.

The distance between the clusters of crystallized nucleosomes (with highly depleted occupancy) scales as 
\begin{equation}
	\frac{L}{N/M}-RM=\left(\frac{W}{\rho}-R\right)\sqrt{\kappa\frac{\rho}{W}}.
\end{equation}

In sum, strongly interacting and well positioned nucleosomes are expected to exhibit highly fluctuating occupancy on large length-scales. 
In Section~\ref{Relevance to empirical results} we discuss how the described above considerations are relevant for more realistic energy landscapes and existing experimental data, but before that we consider effects of another important feature of real systems---autocorrelation of the energy landscape.

\section{Energy landscape with correlations}
\label{Energy landscape with correlations}
So far we discussed random, non-correlated energy landscapes. However, DNA sequence possesses correlations \cite{peng1992long}. Moreover, even on a random DNA sequence an energy landscape is expected to be correlated for distances smaller than $147$ because small shifts of nucleosomes along the DNA does not change completely the sequence covered by the nucleosome. This is why, as we discuss in the next Section, real energy landscape are expected to possess certain autocorrelation. In this Section we discuss how the autocorrelation of the energy landscape affects positioning of nucleosomes. 

We analyze the following scenario in this Section. The energy landscape is assumed to be Gaussian (see Eq. \eqref{Gauss}) with an exponentially decaying autocorrelation, such that
\begin{equation}
	\frac{\left<E_i E_{i+r}\right>}{\sigma^2}=e^{-\frac{r}{r_c}}.
\end{equation}
Here, $r_c \gg 1$ is the correlation distance, such that for distances much larger than $r_c$ the energies are nor correlated, while for distance much smaller than $r_c$ the variation of energy is much smaller than $\sigma$. This model can be mapped to the generalized Random Energy model \cite{derrida1985generalization}. The condition for a good positioning on the single/few bp resolution, we derive below, correspond to the lowest/high temperature phase transition of that model, respectively.

In this Section we assume that for $r_c=0$ the nucleosomes are well positioned on the DNA. The correlation is an additional trouble for positioning and here we derive an additional condition for a good positioning in presence of the autocorrelation, on top of the conditions (\ref{NoIntPos},\ref{GenereicIntCond}), for the non-correlated energy landscapes. We start from the simplest single-nucleosome case.

Consider a single nucleosome on DNA of length $L$ with an Gaussian energy landscape with standard deviation $\sigma$ and exponential autocorrelation with correlation length $r_c \gg 1$. Conceptually we divide the DNA to $L/r_c$ "boxes" of length $r_c$. In order to position the nucleosomes in the box with the highest affinity one needs to satisfy $\sigma \gg \sqrt{2 \ln \frac{L}{r_c}}$. This condition is satisfied because we assume here that without autocorrelation the positioning is good and, therefore, $\sigma \gg \sqrt{2 \ln L}$. Thus, the problem is the positioning of the nucleosome within the box. 

Within the box all the energies are highly correlated. A way to generate such a correlated energy landscape is to set \cite{deserno2010generate}
\begin{equation}
	E_{i+1}=e^{-\frac{1}{r_c}}E_i+\sqrt{1-e^{-\frac{2}{r_c}}} G_i,
\end{equation}
where $G_i$ is an uncorrelated set of Gaussian random variables with standard deviation $\sigma$.
Thus, the standard deviation of energies, $E_i$, if $i$ is in the range much smaller than $r_c$ is given by $\frac{\sigma}{\sqrt{2r_c}}$. With such a standard deviation, to position a nucleosome in a box of size $r_c$, one needs $\frac{\sigma}{\sqrt{2r_c}} \gg \sqrt{2 \ln r_c}$ or
\begin{equation}
	\sigma \gg \sqrt{4 r_c \ln r_c}.
	\label{r_cCond}
\end{equation}  
This is a strong constrain on the positioning. Even for $r_c=5$bp the positioning is bad unless $\sigma$ is much larger than $6k_\text{B}T$. 

Condition \eqref{r_cCond} remains the same also for the case of non-interacting nucleosomes or nucleosomes with only hard-core interactions. This is because \eqref{r_cCond} does not depend on the length of DNA per nucleosome but only on the correlation distance of the energy landscape. This makes positioning of non-interaction (or interacting with only hard-core repulsion) extremely problematic. In the next Section we show that the value of $r_c$ is, at least, tens of base-pairs. For such a correlated energy landscape the positioning condition \eqref{r_cCond} is not expected to be satisfied. 

For interacting nucleosomes the standard deviation of the effective energy landscape is given by  $\sqrt{M}\frac{\sigma}{\sqrt{2r_c}}$, where $M$ is the average number of nucleosomes in a crystallized cluster and given by Eq. \eqref{M}. The positioning condition for strongly interacting nucleosomes is given by
\begin{equation}
	\sigma \gg \sqrt{\frac{4r_c}{M}  \ln r_c} 
	\label{CorrCond1}
\end{equation}
or
\begin{equation}
	\kappa \gg \frac{W}{\rho}\frac{16r_c^2}{\sigma^4}  \ln^2 r_c . 
		\label{CorrCond2}
\end{equation}	

\begin{figure}[h!]
\centering
  \includegraphics[width=0.5\textwidth]{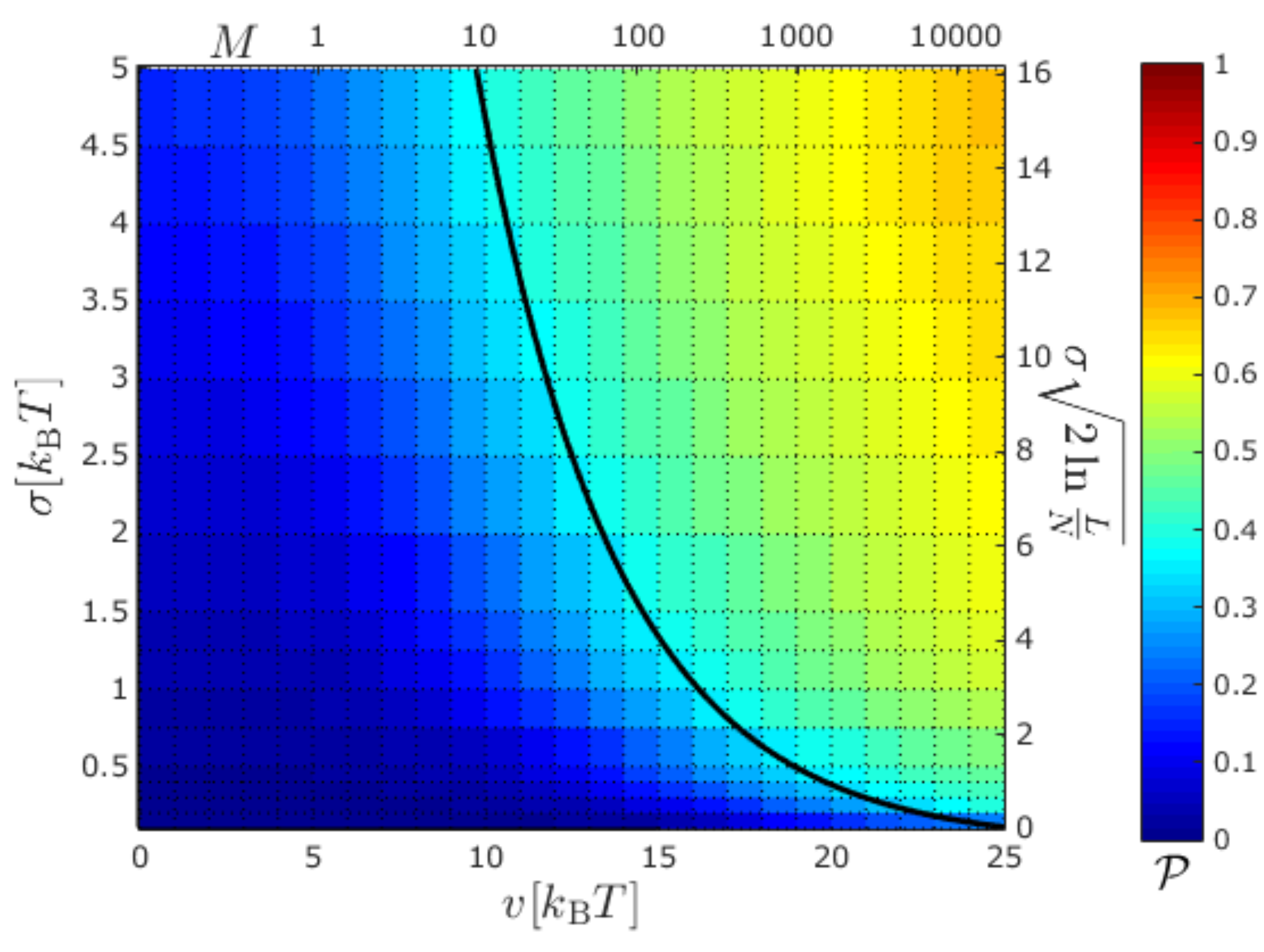}
\caption{Positioning of nucleosomes with interactions on Gaussian energy landscape with an exponentially decaying autocorrelation with correlation coefficient $r_c=20$. Positioning parameter, $\mathcal{P}$ for the DNA with length $L=10^4 \times W$ and $W=147$ for the coverage fraction $\rho=NW/L=80\%$ is plotted vs. disorder strength (left axis) and interaction strength (bottom axis) with preferable distance of $R=148$. On the top one can see the average size of the crystallized cluster of nucleosomes, derived from Eq.~\eqref{M}. On the right the typical binding energy of a nucleosome (relative to the average energy) is shown.
 The line represents the analytic conditions for a good positioning, Eq.~\eqref{CorrCond1} or  \eqref{CorrCond2}.} 
\label{PhaseDiagCorr20}
\end{figure}

\begin{figure}[h!]
\centering
  \includegraphics[width=0.5\textwidth]{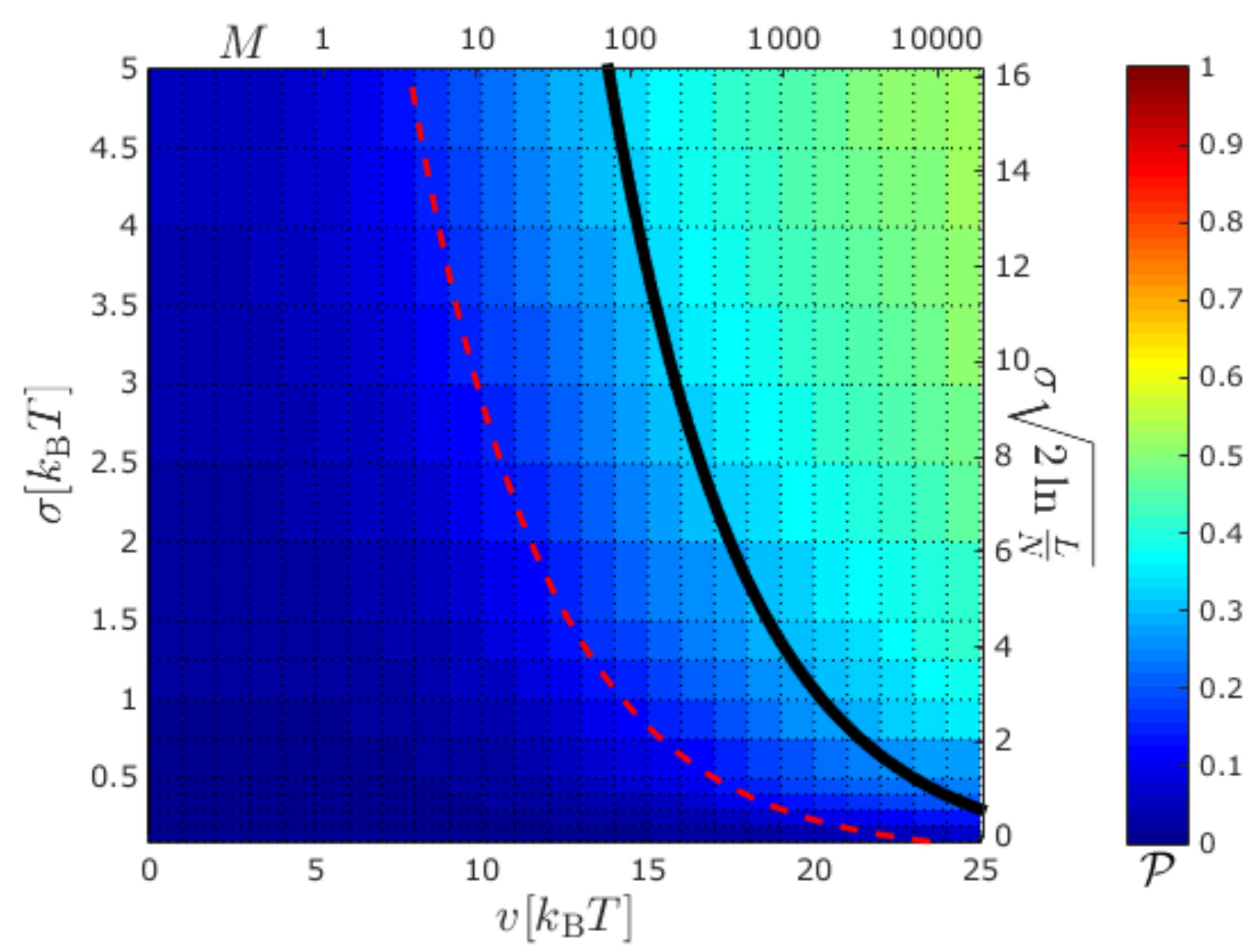}
\caption{Positioning of nucleosomes with interactions on Gaussian energy landscape with an exponentially decaying autocorrelation with correlation coefficient $r_c=100$. Positioning parameter, $\mathcal{P}$ for the DNA with length $L=10^4 \times W$ and $W=147$ for the coverage fraction $\rho=NW/L=80\%$ is plotted vs. disorder strength (left axis) and interaction strength (bottom axis) with preferable distance of $R=148$. On the top one can see the average size of the crystallized cluster of nucleosomes, derived from Eq.~\eqref{M}. On the right the typical binding energy of a nucleosome (relative to the average energy) is shown.
 The solid line represents the analytic conditions for a good positioning on a resolution of $k=1$bp, Eq.~\eqref{CorrCond1} or  \eqref{CorrCond2}. The dahsed line represents the analytic conditions for a good positioning on a resolution of $k=9$bp, Eq.~\eqref{CorrCondk1} or  \eqref{CorrCondk2}.} 
\label{PhaseDiagCorr100}
\end{figure}

In Figs.~\ref{PhaseDiagCorr20} and \ref{PhaseDiagCorr100} one can see the comparison of condition \eqref{CorrCond1} or \eqref{CorrCond2} to numerical results. The obtained results imply that even for short-range autocorrelation of the energy profile with a realistic value of $\sigma$ it impossible to position properly nucleosome on a single bp resolution without strong interactions between them.

\begin{figure*}[!]
\centering
  \includegraphics[width=1\textwidth]{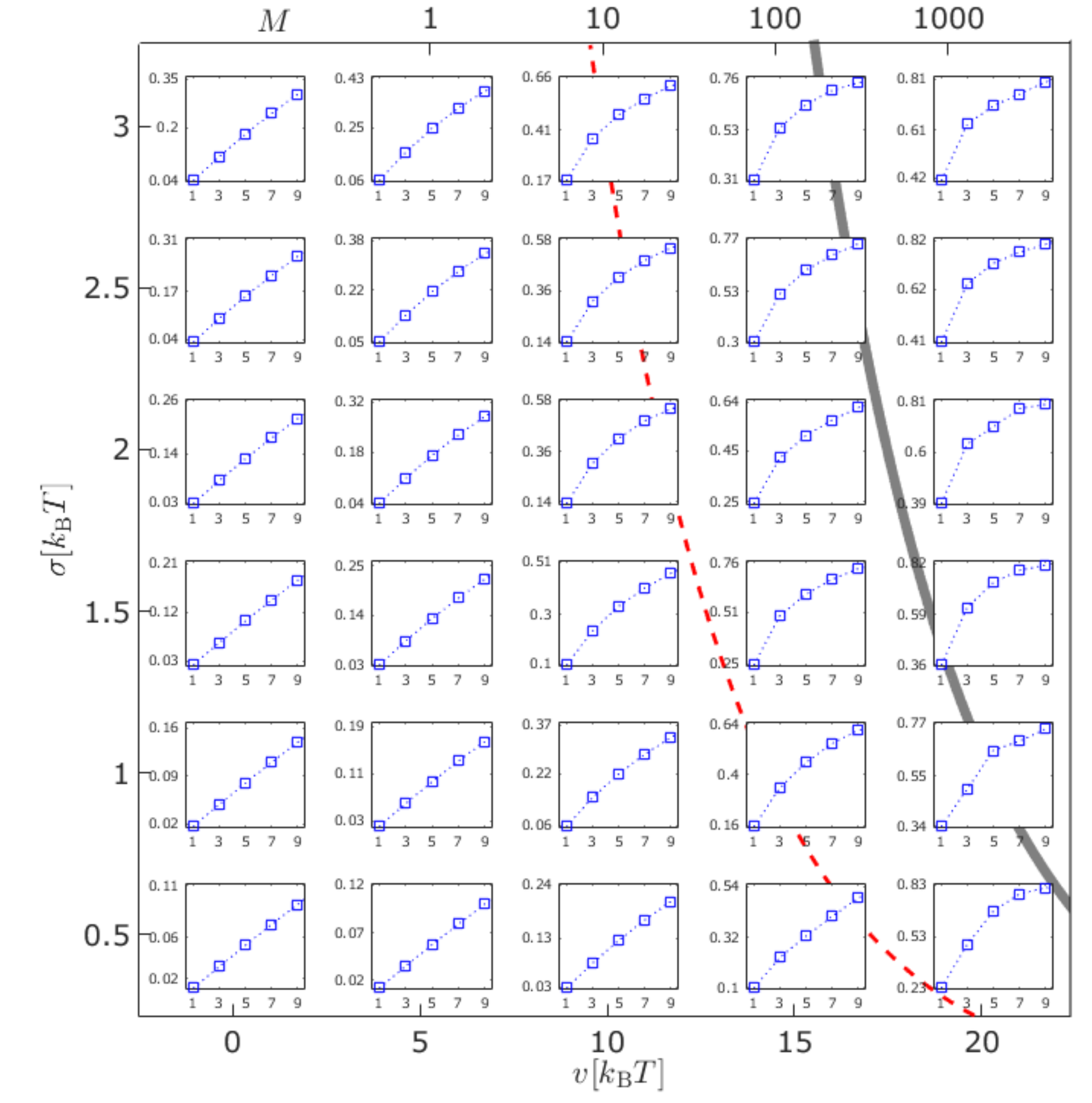}
\caption{Positioning of nucleosomes with interactions on Gaussian energy landscape with an exponentially decaying autocorrelation with correlation coefficient $r_c=100$. Positioning parameter, $\mathcal{P}_k$ for the DNA with length $L=10^4 \times W$ and $W=147$ for the coverage fraction $\rho=NW/L=80\%$ is plotted vs. $k$ for different disorder strengths (left axis) and interaction strengths (bottom axis) with preferable distance of $R=148$. The line represents the analytic conditions for a good positioning on a single bp resoltion, $k=1$, Eq.~\eqref{CorrCond1} or  \eqref{CorrCond2}.} 
\label{PrCorr100}
\end{figure*}

However, the positioning on correlated energy landscape is easier if one allow the nucleosome to be positioned within a few bp. In order to see it we exploit the positioning function, defined in Eq.~\eqref{Pr}. This function, $\mathcal{P}_k$, characterizes the positioning within the resolution of $k$ base-pairs. In Fig.~\ref{PrCorr100} one can see that in some cases even when the single bp resolution positioning is bad, $\mathcal{P}=\mathcal{P}_1 \ll 1$, the positioning within $k=3,5,...$ is significantly better. The condition for $\mathcal{P}_k$ to be of the order one is equivalent to condition \eqref{CorrCond1} or \eqref{CorrCond2} with $r_c$ replaced by $r_c/k$. Namely, the condition for a good positioning within $k$ bp is given by
\begin{equation}
	\sigma \gg \sqrt{\frac{4r_c}{M k}  \ln \frac{r_c}{k}} 
	\label{CorrCondk1}
\end{equation}
or, equivalently,
\begin{equation}
		\kappa \gg \frac{W}{\rho}\frac{16\left(\frac{r_c}{k}\right)^2}{\sigma^4}  \ln^2 \frac{r_c}{k}. 
\label{CorrCondk2}
\end{equation}
One can see in Fig.~\ref{PhaseDiagCorr100} that condition \eqref{CorrCondk1} (or \eqref{CorrCondk2}) can be much weaker than \eqref{CorrCond1} (or \eqref{CorrCond2}). In the next Section we study empirical landscape for which $\sigma$ is roughly $1.5k_\text{B}T$ and $r_c$ is roughly $100$bp. Thus, if the interaction between nucleosomes is strong enough to crystallize them to clusters of size $M = 10-100$ nucleosomes, one would expect to see bad positioning on the level of a single bp resolution with $\mathcal{P}_1=0.1-0.25$ (\eqref{CorrCond1} does not hold) but within $k=9$ bp the nucleosomes are positioned significantly better (\eqref{CorrCondk1} does hold) with $\mathcal{P}_9=0.5-0.8$ (see Fig.~\ref{PrCorr100}). In the next Section we discuss positioning properties on empirical energy landscape and, in general, relevance of the above considerations to real systems.


\section{Relevance to empirical results}
\label{Relevance to empirical results}
So far we discussed positioning on artificial energy landscapes and dissected the phase diagram to different regimes. An obvious question to ask now is: where is the real system on the phase diagram? In this Section we try to get insight into this. To do so, we calculate an energy landscape using a model in Ref.~\cite{kaplan2009dna}. In fact, there are many different models for a binding energy of a nucleosome to a given sequence (examples include \cite{tolstorukov2008nuscore,kaplan2009dna,locke2010high,van2012sequence}, for review see, e.g., \cite{tolkunov2010genomic,teif2012nucleosomes}). We exploit only one of them, from Ref.~\cite{kaplan2009dna}, because we are not interested in predicting locations of nucleosomes on some piece of DNA but in general properties of positioning of nucleosomes. In particular, our goal in this section is to validate our results on artificial energy landscapes and show their relevance to more realistic energy landscapes.

\begin{figure}[h!]
\centering
  \includegraphics[width=0.5\textwidth]{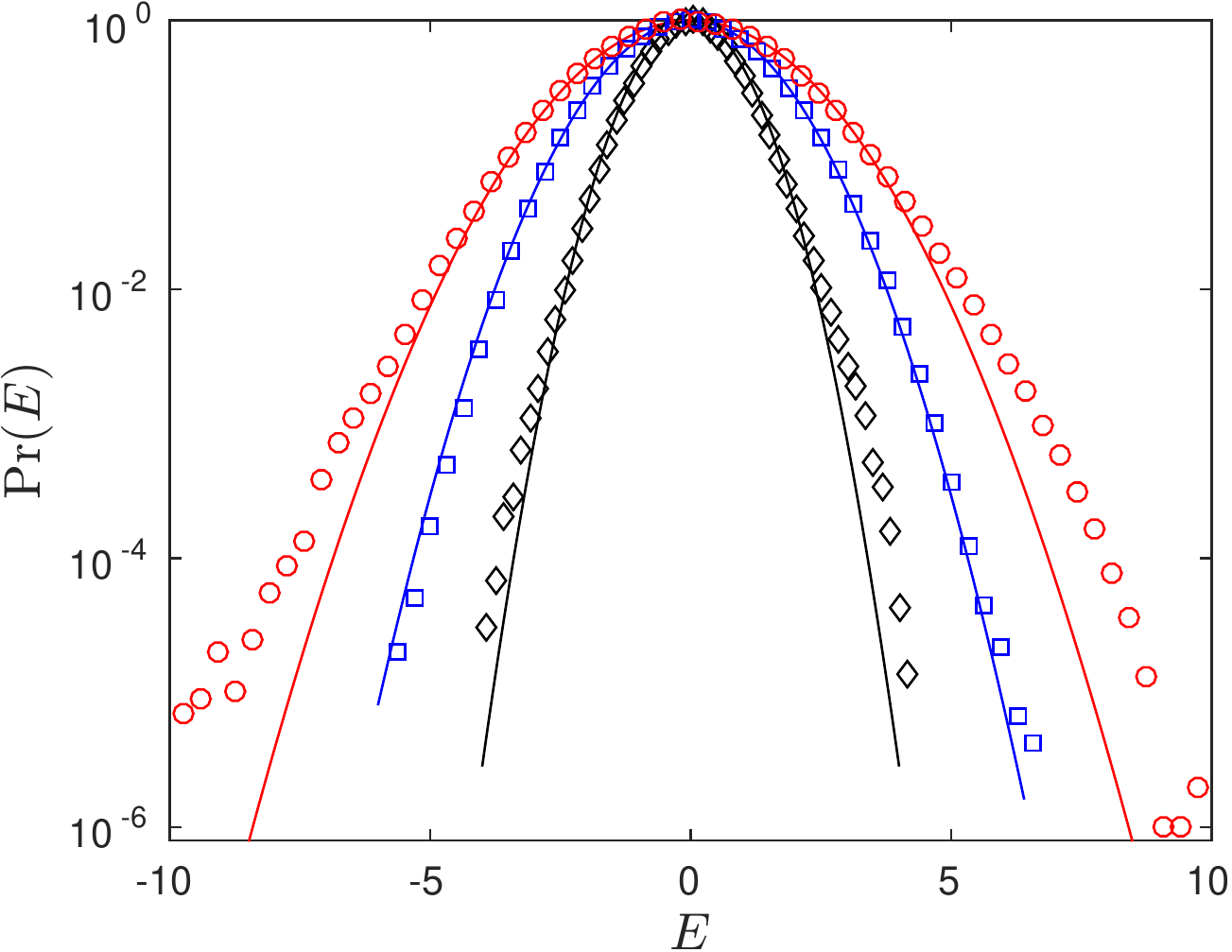}
\caption{Distribution of the binding energy of a nucleosome on a \emph{S. cerevisiae} (circles) and randomly shuffled \emph{S. cerevisiae} genome (squares), calculated using the model in Ref.~\cite{kaplan2009dna}. The standard deviation is equal to $1.6 k_\text{B}T$ for \emph{S. cerevisiae} and $1.24 k_\text{B}T$ for randomly shuffled \emph{S. cerevisiae} genome. The diamonds represent the distribution of the energy landscape calculated in Ref.~\cite{locke2010high} for \emph{S. cerevisiae} genome with a standard deviation of $0.8 k_\text{B}T$. The lines are Gaussian fits with zero mean and respective standard deviations.} 
\label{EnergyDistSegal}
\end{figure}

We start with contrasting the presented, artificial energy landscapes and the one generated using the model in Ref.~\cite{kaplan2009dna} on the genome of \emph{S. cerevisiae}. First thing to note is that the distribution of the binding energies on the \emph{S. cerevisiae} genome is close to, but deviates from a Gaussian (see Fig.~\ref{EnergyDistSegal}). The standard deviation of the energy landscape is given by $1.6k_\text{B}T$. Interestingly, as shown in Fig.~\ref{EnergyDistSegal}, the same model on a randomly shuffled \emph{S. cerevisiae} genome yields significantly narrower energy landscape, well fitted by a Gaussian with $\sigma=1.24k_\text{B}T$. The energy landscape, generated by the model for Ref.~\cite{locke2010high} predicts even narrower energy landscape (also shown in Fig.~\ref{EnergyDistSegal}). 

Energy landscape with such a narrow distribution even with no autocorrelation is not expected to position well the nucleosomes without strong interactions (see Eq. \eqref{NoIntPos} and Fig.~\ref{HardCoreGenericP}). Interaction between nucleosomes can improve the positioning (see Eq. \eqref{GenereicIntCond} and Fig.~\ref{PhaseDiagramNormal}). 

\begin{figure}[h!]
\centering
  \includegraphics[width=0.5\textwidth]{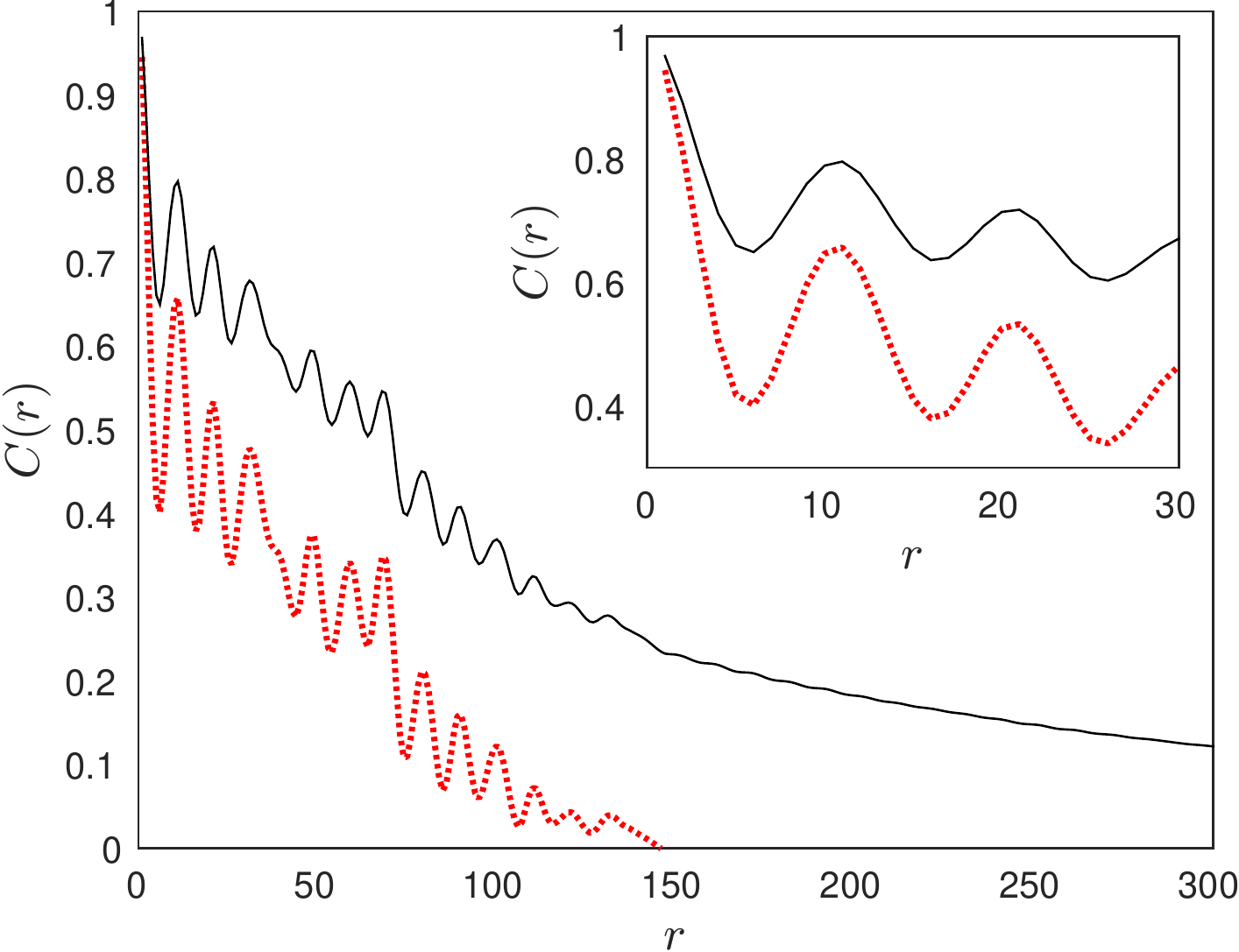}
\caption{Autocorrelation of the binding energy of a nucleosome on a \emph{S. cerevisiae} (black solid line) and randomly shuffled \emph{S. cerevisiae} genome (dotted, red line), calculated using the model in Ref.~\cite{kaplan2009dna}. The inset is a zoom in on the main plot.} 
\label{AutocorrelationSegal}
\end{figure}

However, as is expected, the calculated energy landscape possesses certain autocorrelation. Some part of this autocorrelation is because shifting a nucleosome a few bp doesn't change entirely the bound sequence. This sort of autocorrelation, for distances smaller than $147bp$ exists even on a randomly shuffle genome (see Fig.~\ref{AutocorrelationSegal}). One can clearly see the periodic oscillations with an approximate $10bp$ period \cite{thaastrom1999sequence,trifonov2011cracking}. 

\begin{figure*}[!]
\centering
  \includegraphics[width=\textwidth]{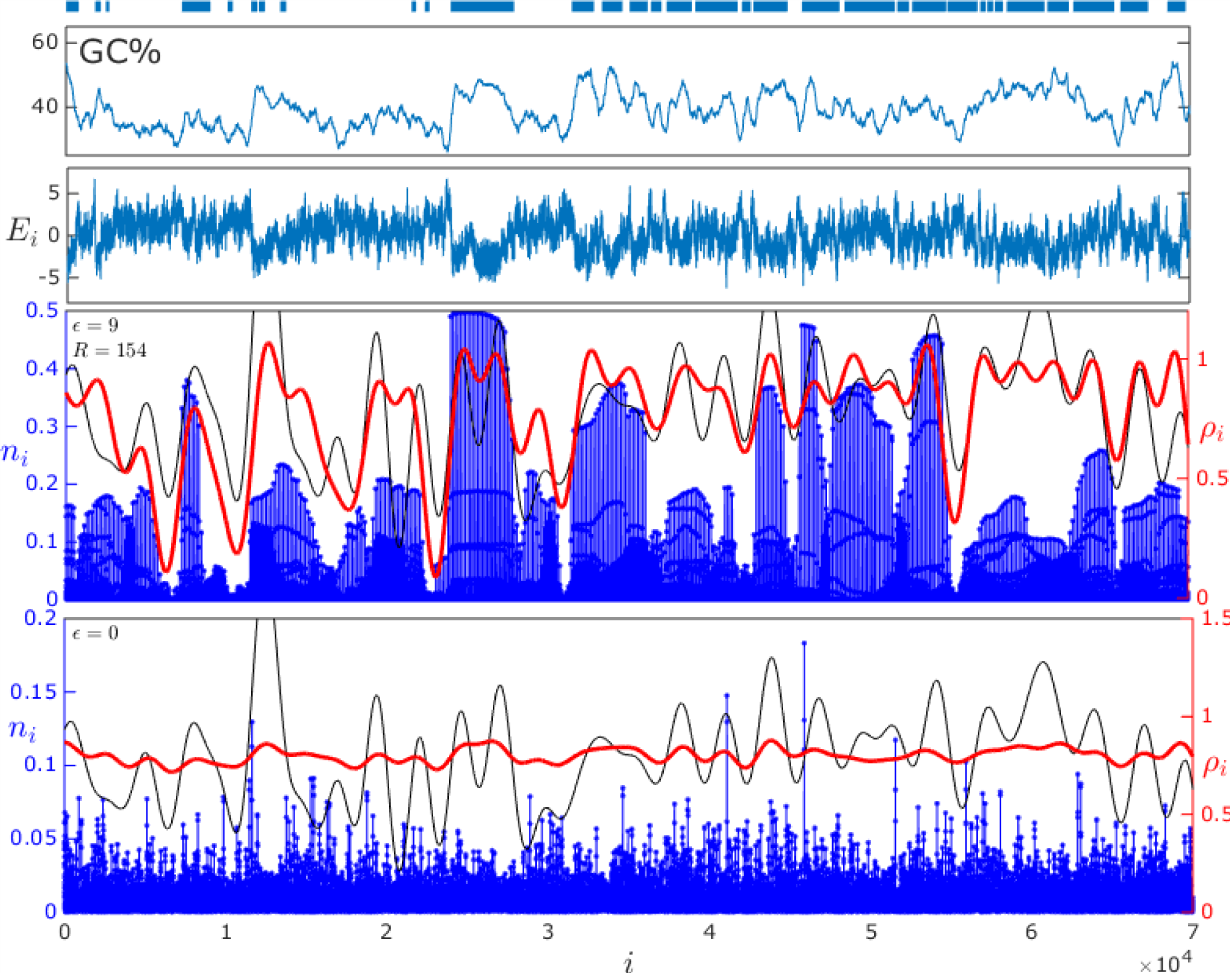}
\caption{Analysis of the first $7 \cdot 10^4 bp$ of \emph{S. cerevisiae}'s first chromosome. The lines on top of the figure depict exons. In panel number (I) the GC content is plotted. (II) Binding energy profile calculated using model from Ref.~\cite{kaplan2009dna}. The additive constant term is such that the mean energy along the chromosome is zero. (II) Calculation of the nucleosome distribution for interacting nucleosomes with $R=154$ and $ v =9$. The chemical potential is such that $\rho=0.8$. The lines with dots represent $n_i$ values, while the thick, red line represent occupancy level. The thin black line is the occupancy from the data in Ref.~\cite{brogaard2012map}. Both the occupancies are smoothed, such that only $100$ lowest Fourier modes are presented. (III) The same as for the (II) panel but with only hard-core interactions, $ v =0$.} 
\label{OccupancySegalStartChr1}
\end{figure*}

\begin{figure*}[!]
\centering
  \includegraphics[width=\textwidth]{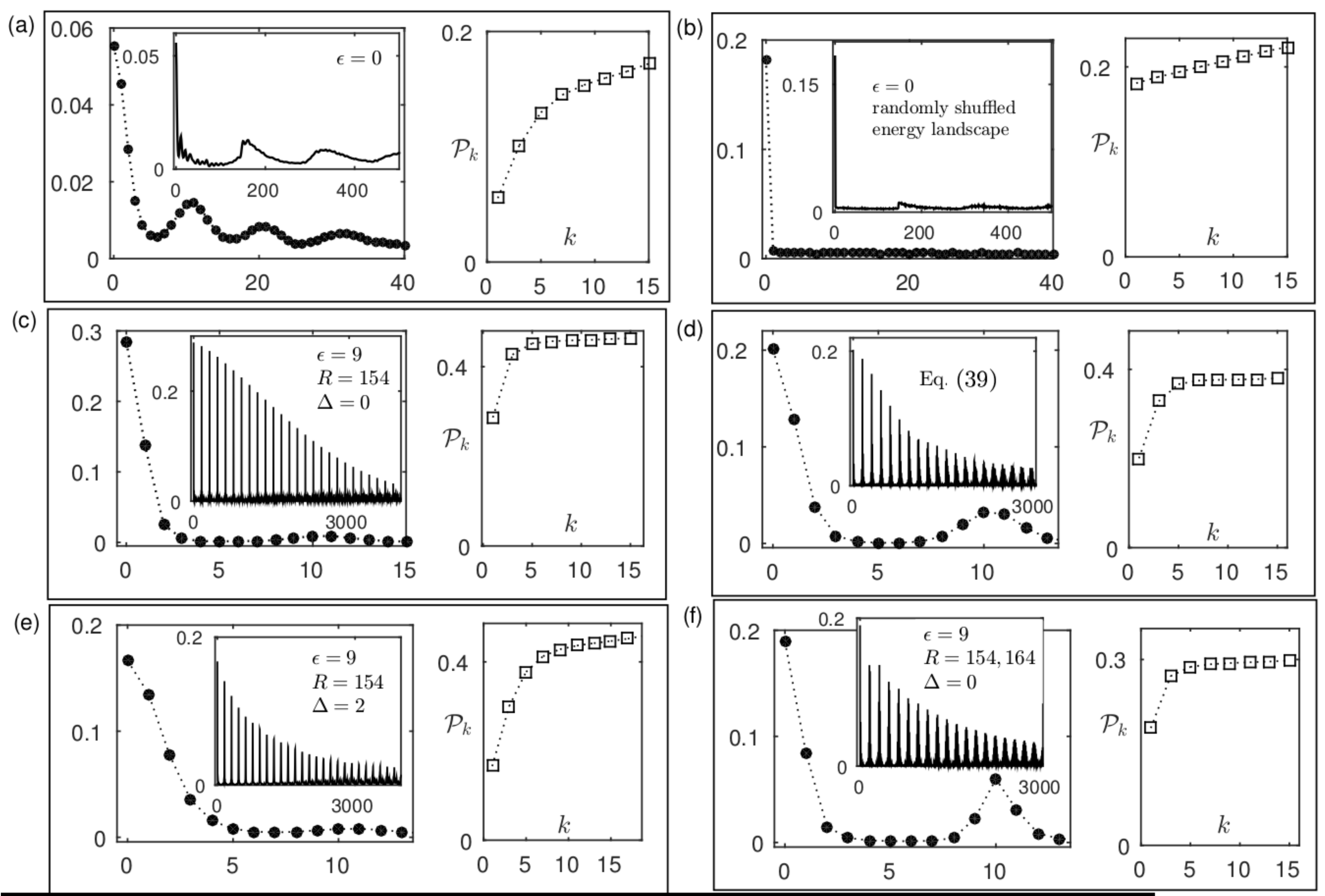}
  \put(-145,199){\fcolorbox{white}{white}{(39)}}
\caption{Positioning goodness in different cases. In each panel the left plots are the average profile of the $N$ largest values of $n_i$ (insets are zoom out of the main plots). The right plots are the positioning function $\mathcal{P}_k$ vs. $k$. (a) Energy landscape from Ref.~\cite{kaplan2009dna} with only hard-hore interactions between the nucleosomes, $v=0$. (b) The same as in (a) but the energy landscape (not the sequence) is randomly shuffled. (c) The same as in (a) but nucleosomes interact with $v=9$, $R=154$, $\Delta=0$. (d) The same as in (a) but nucleosomes interact with the potential from Eq.~\eqref{35}. (e) The same as in (c) but nucleosomes interact with $\Delta=2$. (f) The same as in (a) but nucleosomes interact with two wells potential, $R=154$ and $R=164$.} 
\label{PosParFunction}
\end{figure*}

On top of that, due to some sequence correlation along the real, non-shuffled, \emph{S. cerevisiae} genome, the autocorrelation of binding energy persist even for distances larger than $147bp$ (see Fig.~\ref{AutocorrelationSegal}). One reason for such a long-scale autocorrelation is that model in Ref.~\cite{kaplan2009dna} is biased by GC content \cite{chung2010effect} and GC content possess significant autocorrelation along the \emph{S. cerevisiae} genome \cite{bradnam1999g} (see Fig.~\ref{OccupancySegalStartChr1}). The autocorrelation function clearly deviates from the simple exponential decay, which we assumed in our theoretical considerations above. Roughly, the correlation distance is close to $100$bp, making the positioning much more problematic, relative to uncorrelated energy landscape with the same standard deviation (see Fig.~\ref{PosParFunction}(a) vs. (b)).

On such an auto-correlated energy landscape, with only hardcore interactions, $ v =0$, nucleosomes are not well positioned on a single bp resolution (see a typical $n_i$ profile in Fig.~\ref{OccupancySegalStartChr1}). Namely, the single bp positioning parameter, $\mathcal{P}=0.06$ (calculated on the first chromosome of \emph{S. cerevisiae}), is much smaller than $1$. With only hard-core interactions the peaks in $n_i$ are not only low, but also wide. In fact the positioning function $\mathcal{P}_k$, shown in Fig.~\ref{PosParFunction}(a), demonstrates that the nucleosomes are "fuzzy" and poorly positioned even on the resolution of $10bp$.

The absence of good positioning can be partially attributed to strong autocorrelation of energy, because the distribution of uncorrelated Gaussian energy landscape with $\sigma=1.6k_\text{B}T$ results in much better positioning, as is discussed in Section \ref{No Interaction Generically disordered energy landscape}. Indeed, calculation of positioning of nucleosomes with only hard-core interaction on randomly shuffled energy landscape, calculated using model in Ref.~\cite{kaplan2009dna} results in narrow ($1bp$) peaks of $n_i$ of an average height of $\mathcal{P}_1=\mathcal{P}=0.18$ (see Fig.~\ref{PosParFunction}(b)).

In contrast to the case with only hard-core interactions, the nucleosomes can be positioned much better in presence of interactions between neighboring nucleosomes. For example, for the interaction strength of $ v =9$ and the preferable distance $R=154$ the peaks of $n_i$ are much higher, such that the positioning parameter is $\mathcal{P}_1=0.3$ (see Figs.~\ref{OccupancySegalStartChr1} and \ref{PosParFunction}(c)). The width of the peaks is $3-5bp$,  such that the positioning function is $\mathcal{P}_k \sim 0.45$ for $k \geq 3$, as can be seen in Fig.~\ref{PosParFunction}(c). 

Can one position nucleosome with a more realistic interaction potential? A reasonable choice seems to be the one used in Ref.~\cite{chereji2011statistical2} to fit qualitatively the $10n+5$ (or, sometimes, $10.6n+8$ \cite{cohanim2006three}) periodicity found in many works, starting from Ref.~\cite{lohr1979organization}. In order to verify that our results do not depend qualitatively on the precise form of the interaction potential we used the same form as in Ref.~\cite{chereji2011statistical2}, but with higher prefactor ($12$ instead of $5$) and with a cutoff of $180$bp for the computational purposes:
\begin{equation}
	V(r)=
	\begin{cases} 
	\infty &  r<W  \\
	12k_\text{B}T\cos\left(\frac{2 \pi(r-W)}{10\text{bp}} \right)e^{-\frac{r-W}{50\text{bp}}} & W \leq r \leq 180 \\
	0 & r >180
	\end{cases}.
	\label{35}
\end{equation}

In Fig.~\ref{PosParFunction}(d) one can see that the interaction potential in Eq.~\eqref{35} is able to position nucleosomes on the energy landscape generated using the model in Ref.~\cite{kaplan2009dna} within the resolution of $3-5$bp. In the next Section \ref{Tunability and robustness of the positioning} we analyze in more detail robustness and tunability of positioning to different properties of the interacting potential.

In sum, these results indicate that good positioning of nucleosomes is possible even on a realistic energy landscapes with narrowly distributed (small $\sigma$) and highly auto-correlated (large $r_c$) energies, provided strong enough interactions between them. 

Beyond the positioning of nucleosomes on small length scales, as we described in Section \ref{Large-scale fluctuations of occupancy}, strong interactions between neighboring nucleosomes change DNA occupancy by nucleosomes on large length scales. We turn now to discuss how the theoretical predictions in Section \ref{Large-scale fluctuations of occupancy} are relevant for realistic scenarios.

\begin{figure*}[!]
\centering
  \includegraphics[width=\textwidth]{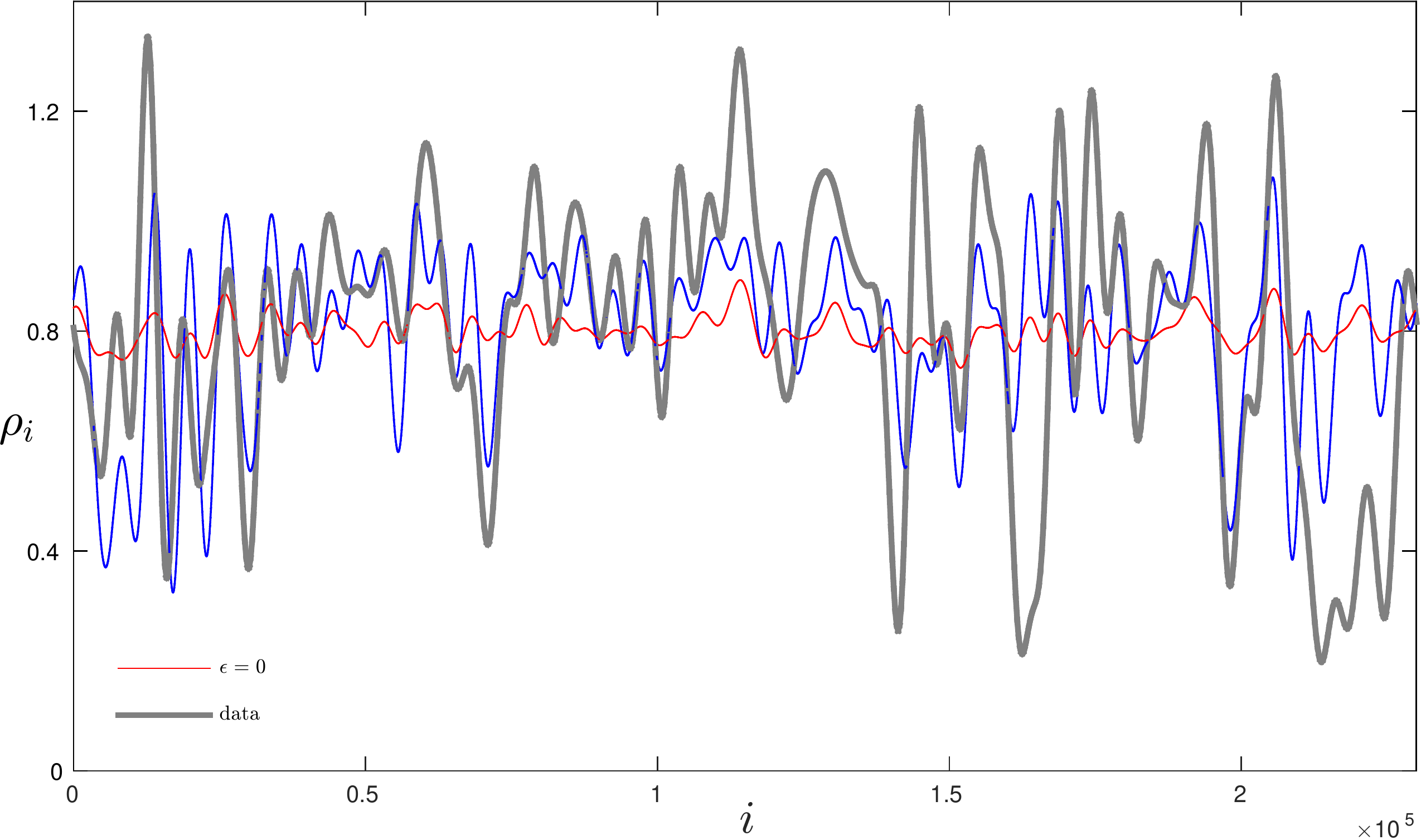}
\caption{Occupancy of nucleosomes on the global scale of the whole first chromosome of \emph{S. cerevisiae}. Thin blue line represents calculation of the nucleosome distribution for interacting nucleosomes with $R=154$ and $ v =9$. Thick, gray line is the occupancy from the data in Ref.~\cite{brogaard2012map}. The calculated occupancy for nucleosomes with only hard-core interactions is represented by the thin, red, almost flat line. The chemical potential set for calculations is such that $\rho=0.8$. All the occupancies are smoothed, such that only $50$ lowest Fourier modes are presented.} 
\label{OccupancySegal}
\end{figure*}

\subsection{Large-scale fluctuations of occupancy}
We start with nucleosomes with only hard-core interactions, as a reference case. In this case, since the energy profile is not well correlated for long distances, the occupancy does not fluctuate significantly on large length scales (above a few nucleosome repeat lengths), as how in Figs.~\ref{OccupancySegalStartChr1} and \ref{OccupancySegal}. However, as discussed in Section \ref{Large-scale fluctuations of occupancy}, interactions locally crystallize nucleosomes, inducing large scale fluctuations in occupancy. In Figs.~\ref{OccupancySegalStartChr1} (zoom on first $7 \cdot 10^4$ bp of chromosome I of \emph{S. cerevisiae}) and \ref{OccupancySegal} (whole chromosome I of \emph{S. cerevisiae}) one can see that interacting nucleosomes are distributed nonuniformly along the DNA, in contrast to nucleosomes with only hard-core interactions. Interestingly, this sort of large length scale nonuniformity one also observes on the single-bp resolution data \cite{brogaard2012map} (see Figs.~\ref{OccupancySegalStartChr1} and \ref{OccupancySegal}). Moreover, the calculated occupancy seems to follows quite consistently the experimental one on different length scales. This is especially surprising because the data from \cite{brogaard2012map} is based on chemical cleavage, while the model used by use to calculate energy profile along the DNA was derived based on MNase digestion \cite{kaplan2009dna}. 

These results indicates that the interactions between the nucleosomes help to position them not only on the short length-scales (a few bp), but also on the long length-scales, inducing large long-scale fluctuations in nucleosomes occupancy. In other words, strong interactions between the nucleosomes naturally yields long nucleosomes diluted and enriched regions along the genome.

We summarize this Section with the following conclusions.
On a realistic energy landscape nucleosomes are much better localized in presence of strong interactions. In this case the calculated results, predicting large length scale occupancy fluctuations, agree qualitatively and, surprisingly, quantitatively with the experimental data. We turn now to a more detailed analysis of how properties of the interaction potential affect distribution of nucleosomes along the genome.

\section{Tunability and robustness of the positioning}
\label{Tunability and robustness of the positioning}
Several relevant question are still to be answered. How robust are the obtained results for the made assumptions? What if the interaction potential possesses a certain width ($\Delta >1$ in Eq.~\eqref{4})? How robust the positions of nucleosomes to change of parameters, like the strength of the interactions $ v $, preferable distance, $R$, number of wells, and the width of the interactions potential, $\Delta$. Can one "tune" the distribution of nucleosomes along the DNA changing (locally or globally) the parameters of the interaction potential? 

We addressed some of these issues above, considering more realistic potential between the nucleosomes, in Eq. \eqref{35}, and found that the conclusions are quite robust to a particular form of the potential. In this Section we address these questions more systematically. We focus here on the energy profile from the previous Section, calculated using model from Ref.~\cite{kaplan2009dna} and take as a starting point potential with a single well with $R=154$, $v=9$, $\Delta=0$. In this case the positioning is reasonably good (see Figs.~\ref{OccupancySegalStartChr1} and \ref{PosParFunction}(c)). 

\begin{figure*}[!]
\centering
  \includegraphics[width=\textwidth]{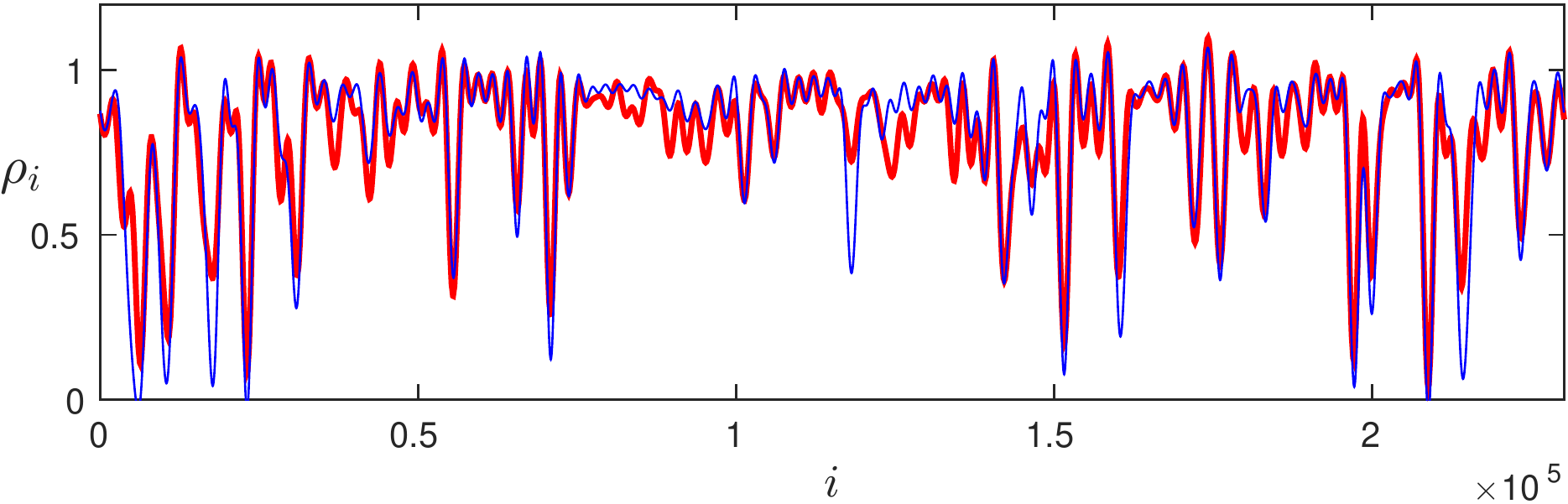}
\caption{Changing the width of the potential, $\Delta$. Thick red line represents calculation of the nucleosome distribution for interacting nucleosomes with $R=154$, $ v =9$ and $\Delta=0$. Thin, blue line represents calculation of the nucleosome distribution for interacting nucleosomes with $R=154$, $ v =9$ and $\Delta=2$, such that the width of the potential well is $2 \Delta+1=5bp$. The chemical potential set for calculations is such that $\rho=0.8$. All the occupancies are smoothed, such that only $100$ lowest Fourier modes are presented. Locally the occupancy values are also highly correlated. The Pearson correlation coefficient of $\rho_i$ (without any smoothing) for $\Delta=0$ and $2$ is $0.87$. The values of $\rho_i$ for $\Delta=0$ and $2$ differ by $19\%$ on average. However, the locations of nucleosomes for $\Delta=0$ and $2$ significantly differ: $N=\rho L/W$ locations with the highest values of $n_i$ for the two cases possess an overlap of only $18\%$.} 
\label{TuneDeltaGlobal}
\end{figure*}

We start by changing the width of the interaction potential, $\Delta$. Before we always (except from analyzing the potential form from Eq. \eqref{35}) assumed that the interaction potential is sharp to the level of a single bp, $\Delta=0$. Of course this assumption is not realistic and real effective interaction potentials probably possess a finite width and more than one energy well \cite{strauss1983organization,widom1992relationship,chereji2011statistical}. However, changing the width to a few base-pairs does not change qualitatively the results. Taking $\Delta=2$ the positioning remains good on the length scale of $2\Delta+1=5$, such that the typical peak of $n_i$ has a height of $\mathcal{P} \simeq 0.3$ and width of $5$bp, as shown in Fig.~\ref{PosParFunction}(e). 
As one can see in Fig.~\ref{TuneDeltaGlobal}, the occupancy on large length scales does not change much as one tune the value of $\Delta$. Even on the level of a single bp resolution the Pearson correlation coefficient of $\rho_i$ for a potential with a width of one bp ($\Delta=0$) and $5$bp ($\Delta=2$) is $0.87$. The precise locations of nucleosomes do change, however. Looking on the $N=\rho L/W$ largest values of $n_i$ for both values of $\Delta$ we observe an overlap of $18\%$. This value makes sense because we expect that, upon changing the width of the potential well from $1$ bp to $5$bp, $20\%$ of the nucleosomes will remain in their positions and $80\%$ will move $\pm 2$ bp, within the new potential well. In sum, widening of the potential width does not change the distribution of nucleosomes on a large scale, but makes their position uncertain on the length scale of $2\Delta+1$.

\begin{figure*}[!]
\centering
  \includegraphics[width=\textwidth]{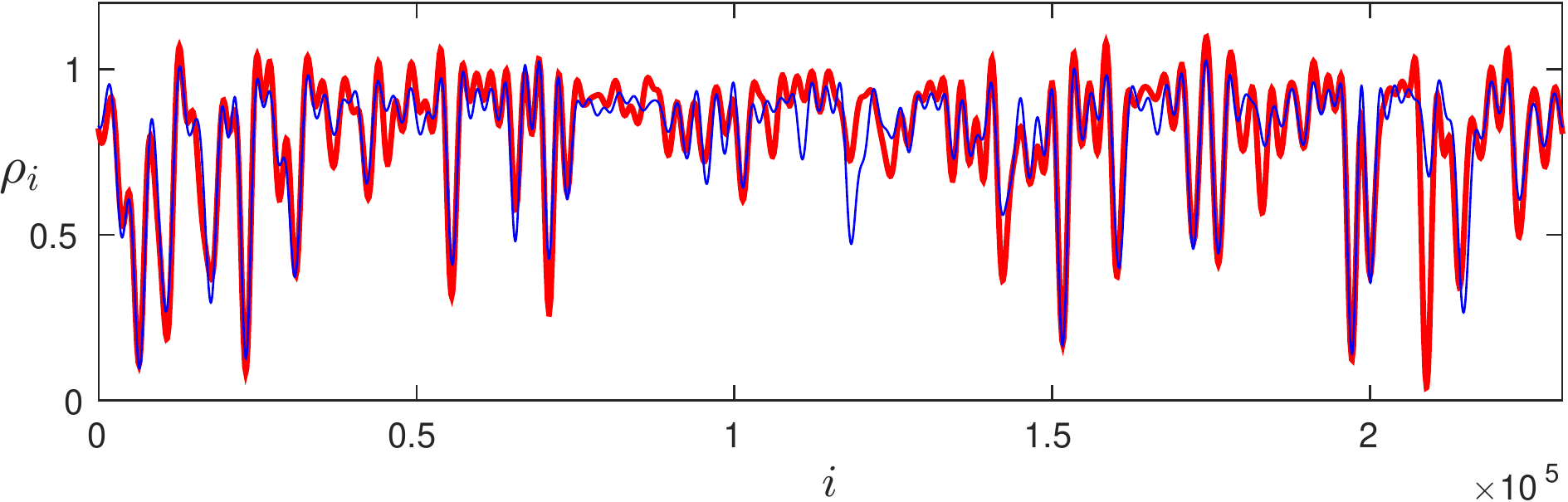}
\caption{Changing the number of wells of the potential. Thick red line represents calculation of the nucleosome distribution for interacting nucleosomes with $R=154$, $ v =9$ and $\Delta=0$. Thin, blue line represents calculation of the nucleosome distribution for interacting nucleosomes with two wells (both with $\Delta=0$) $R=154$ and $R=164$, $ v =9$. The chemical potential set for calculations is such that $\rho=0.8$. All the occupancies are smoothed, such that only $100$ lowest Fourier modes are presented. Locally the occupancy values are also highly correlated. The Pearson correlation coefficient of $\rho_i$ (without any smoothing) for one and two wells is $0.75$. The values of $\rho_i$ for one and two wells differ by $18\%$ on average. However, the locations of nucleosomes for one and two wells significantly differ: $N=\rho L/W$ locations with the highest values of $n_i$ for the two cases possess an overlap of only $3\%$.} 
\label{TuneSecondWellGlobal}
\end{figure*}

We add now one more well to the interaction potential at $10$bp from the first one, $R=154+10=164$. The positioning gets slightly worse, such that the typical peak of $n_i$ has a height of $\mathcal{P} \simeq 0.2$ and width of $3$bp, as shown in Fig.~\ref{PosParFunction}(f). 
As one can see in Fig.~\ref{TuneSecondWellGlobal}, the occupancy on large length scales does not change much as one tune the value of $\Delta$. Even on the level of a single bp resolution the Pearson correlation coefficient of $\rho_i$ for a potential with one and two wells is $0.75$. The precise locations of nucleosomes do change, dramatically, however. Looking on the $N=\rho L/W$ largest values of $n_i$ for both values of $\Delta$ we observe an overlap of only $3\%$. In sum, another well in the potential width does not change the distribution of nucleosomes on a large scale, but does change their positions on small length scale.

\begin{figure*}[!]
\centering
  \includegraphics[width=\textwidth]{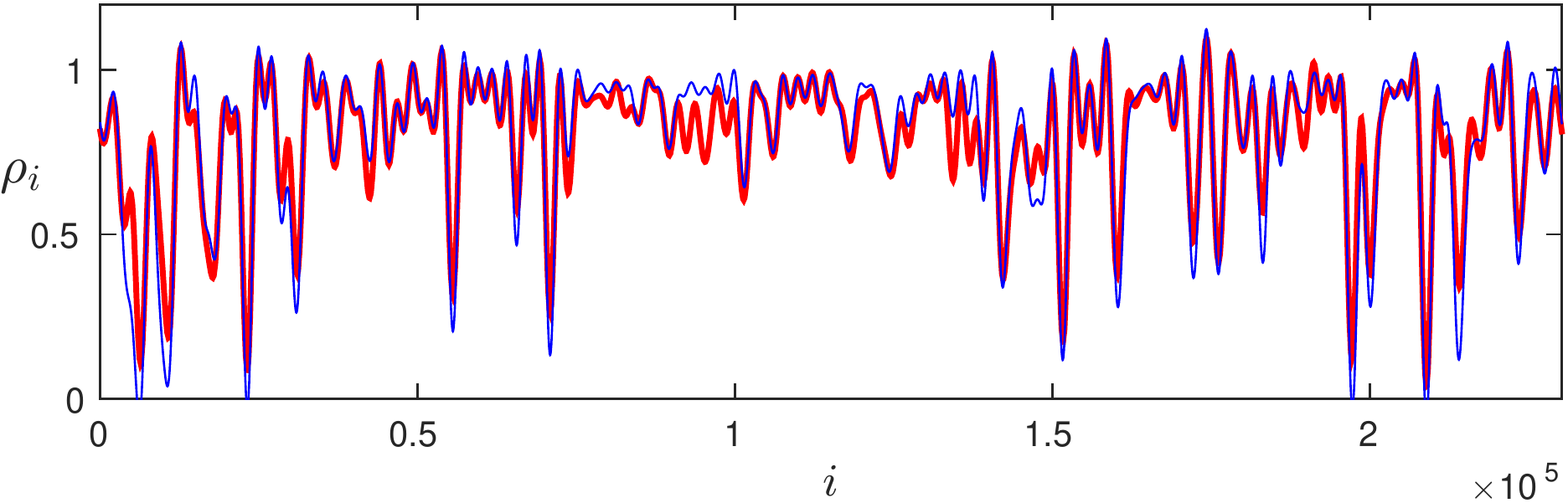}
\caption{Changing the strength of the potential, $ v $. Thick red line represents calculation of the nucleosome distribution for interacting nucleosomes with $R=154$, $ v =9$ and $\Delta=0$. Thin, blue line represents calculation of the nucleosome distribution for interacting nucleosomes with $R=154$, $ v =11$ and $\Delta=0$. The chemical potential set for calculations is such that $\rho=0.8$. All the occupancies are smoothed, such that only $100$ lowest Fourier modes are presented. Locally the occupancy values are also highly correlated. The Pearson correlation coefficient of $\rho_i$ (without any smoothing) for $ v =9$ and $11$ is $0.94$. The values of $\rho_i$ for $ v =9$ and $11$  differ by $15\%$ on average. In this case even the locations of nucleosomes for $ v =9$ and $11$ do not significantly differ: $N=\rho L/W$ locations with the highest values of $n_i$ for the two cases possess an overlap of $67\%$.} 
\label{TuneEpsGlobal}
\end{figure*}

The depth of the potential well does affect the goodness of positioning and has to be strong enough to have any effect. However, once a reasonable positioning is achieved, its value does not change things much. As one can see in Fig.~\ref{TuneEpsGlobal}, the occupancy on a large length scales does not change much as one tune the value of $ v $ from $9$ to $11$. Even on the level of a single bp resolution the Pearson correlation coefficient of $\rho_i$ for a potential with one and two wells is $0.94$. Even the precise locations of nucleosomes do not change, dramatically. Looking on the $N=\rho L/W$ largest values of $n_i$ for both values of $v$ we observe an overlap of $67\%$.

\begin{figure*}[!]
\centering
  \includegraphics[width=\textwidth]{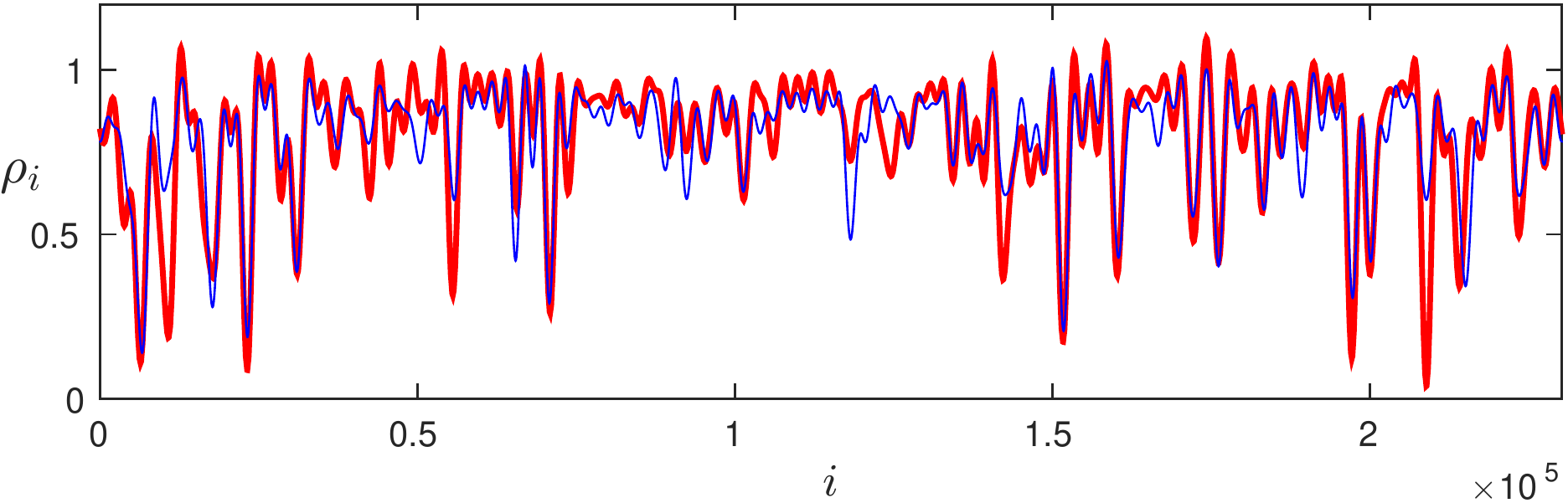}
\caption{Changing the well location of the potential, $R$. Thick red line represents calculation of the nucleosome distribution for interacting nucleosomes with $R=154$, $ v =9$ and $\Delta=0$. Thin, blue line represents calculation of the nucleosome distribution for interacting nucleosomes with $R=160$, $ v =9$ and $\Delta=0$. The chemical potential set for calculations is such that $\rho=0.8$. All the occupancies are smoothed, such that only $100$ lowest Fourier modes are presented. Locally the occupancy values are also highly correlated. The Pearson correlation coefficient of $\rho_i$ (without any smoothing) for $R=154$ and $160$ is $0.62$. The values of $\rho_i$ for $R=154$ and $160$  differ by $23\%$ on average. On the resolution of a few bp the locations of nucleosomes for $R=154$ and $160$ are very different: $N=\rho L/W$ locations with the highest values of $n_i$ for the two cases possess an overlap of only $1\%$.} 
\label{TuneRglobal}
\end{figure*}

As one would expect, the position of the well, $R$, does not affect significantly the positioning properties and distribution of the nucleosomes on a large scale. As one can see in Fig.~\ref{TuneRglobal}, the occupancy on large length scales does not change much as one tune the value of $R$ from $154$ to $160$. On the level of a single bp resolution the Pearson correlation coefficient of $\rho_i$ for a potential with one and two wells is $0.62$. However, precise positioning of nucleosomes is very sensitive to the value of $R$:  $N=\rho L/W$ locations with the highest values of $n_i$ for the two cases of $R=154$ and $R=160$ possess an overlap of only $1\%$. 

We conclude that, in principle, a cell, tuning the preferable distance between nucleosomes, can control their distribution along some part of DNA without changing significantly large scale properties, like an average positioning goodness and large length scale occupancy.

\section{Summary}
\label{Summary}
In this article we focus on goodness of positioning of nucleosomes on the DNA. We make several simplifying assumptions. We assume that we can take into account all the positioning factors by have an effective energy landscape and an effective interaction potential between neighboring nucleosomes. We ignore that nucleosomes can invade each others DNA territories. In addition we analyze only the equilibrium distribution of nucleosomes, ignoring very probable non-equilibrium aspects of nucleosomes positioning. However, even within this simplified framework we clarify a few aspects of nucleosome positioning, which do not seem to depend on these details.

Looking on a generic energy landscape with some energy distribution width, some energy typical autocorrelation distance and some interaction potential between neighboring nucleosome we derive condition for a good positioning. We briefly summarize the conditions in the following paragraph.

Assuming that neighboring nucleosomes possess a preferable distance with an affinity $\kappa \gg 1$, relative to other distances, the number of locally crystallized nucleosomes, $M \gg 1$, is given by Eq.~\eqref{M}, where $N$ is the average number of nucleosomes and $L$ is the length of the DNA. The the positioning is expected to be good on an uncorrelated Gaussian disorder with standard deviation $\sigma$ is condition Eq.~\eqref{srCond} holds.
On a correlated energy landscape with a certain correlation distance, $r_c$, the positioning conditioning depends on the required resolution. For a good positioning within $k$ bp it is given by Eq.~\eqref{CorrCondk1}.

Importantly, without strong interactions, $\kappa \simeq 1$, the conditions do not seem to hold for realistic parameters, indicating an important role of effective interactions between the nucleosomes in their positioning. 
If the positioning, as we suggest, is controlled by interactions, one expect to see long length-scale fluctuations of nucleosome occupancy. This conclusion agrees with empirical data on occupancy of nucleosomes. Moreover, the derived large length-scale occupancy profile, derived from an effective energy landscape and interaction potential which is sufficient to position the nucleosomes is similar to the empirical one. We also analyze the robustness of positioning to parameters of the model. The parameters can vary with time and be different on different parts of the genome. In fact, as we demonstrate, tuning some parameters one can dramatically change the distribution of nucleosomes. 
In sum, our study emphasizes an important role of interaction between the nucleosomes and indicates range of parameters needed for it. We expect this knowledge to be important for better understanding of organization of our epigenome.

\appendix
\section{Non-Gaussian energy landscapes}
The distribution of the binding energies along the DNA does not have to be Gaussian. Here we analyze two other possible scenarios for the energy landscapes.
\subsection{Positioning of one nucleosome}
\label{Appendix Positioning of one nucleosome}
Here we discuss positioning of a single nucleosome on up-exponential and down-exponential energy landscapes.
\subsubsection{Disordered energy landscape with \emph{down}-exponential distribution}
Consider scenario with what we denote as \emph{down}-exponential distribution:
\begin{equation}
	\Pr(E_i)=\frac{1}{\mathcal{E}}e^{\frac{E_i}{\mathcal{E}}} \ ; \ -\infty < E_i \leq 0 \ ; \ \mathcal{E}>0.
	\label{downexp}
\end{equation} 
In this case the typical minimal energy can be estimated using Eq.~\eqref{E1eq} to be
\begin{equation}
	E_1^\text{o} \simeq -\mathcal{E} \ln L.
	\label{E1downexp}
\end{equation}
The partition function be be estimated separately in two regimes (two phases in the thermodynamic limit): 
in the nonfrozen regime, $\mathcal{E} \ll 1$ the partition function is given by
\begin{equation}
	Z \simeq L \int_{-\infty}^0 e^{-E}\Pr(E)dE=\frac{L}{1-\mathcal{E}}.
	\label{Zdownexp}
\end{equation}
The positioning parameter is this case is given by
\begin{equation}
	\mathcal{P} \simeq L^{\mathcal{E}-1} \ll 1,
\end{equation}
such that the positionig is poor for $\mathcal{E} \ll 1$.

In the opposite regime $\mathcal{E} \gg 1$ the integral in Eq.~\eqref{Zdownexp} diverges and the partition function is dominated by the deepest wells. Thus, it can be estimated using $k$-minimal energies, given by
\begin{equation}
	E_k^\text{o} \simeq \mathcal{E} \ln \frac{L}{k}.
\end{equation}
Therefore, 
\begin{equation}
	Z \simeq \sum_{k=1}^{\infty}\left(\frac{L}{k}\right)^{\mathcal{E}} = L^\mathcal{E}\zeta(\mathcal{E}),
\end{equation}
such that the positioning parameter in this regime is given by
\begin{equation}
	\mathcal{P} \simeq \frac{1}{\zeta(\mathcal{E})}.
\end{equation}
and is larger than $1/2$ and close to one for 
\begin{equation}
	\mathcal{E} \gg \zeta^{-1}(2) \simeq 1.7.
	\label{ExpDownCond1}
\end{equation}

In sum, for down-exponential distribution of energies the only requirement for a good positioning is that the parameter $\mathcal{E}$ is larger than one for any length of the DNA. 

\subsubsection{Disordered energy landscape with \emph{up}-exponential distribution}
Consider another scenario with what we denote as \emph{down}-exponential distribution:
\begin{equation}
	\Pr(E_i)=\frac{1}{\mathcal{E}}e^{-\frac{E_i}{\mathcal{E}}} \ ; \ 0 \leq E_i < \infty  \ ; \ \mathcal{E}>0.
	\label{upexp}
\end{equation} 
In this case there are many energies which are close to zero, $E_1^\text{o} \simeq E_2^\text{o}\simeq E_3^\text{o}\simeq... \simeq 0$. Positiong is impossible for such energy landscape; for any value of $\mathcal{E}$ one has $\mathcal{P} \ll 1$. However, as we show below interaction between nucleosomes can induce reasonable positioning even on such a energy landscape.

\subsection{Positioning of multiple nucleosomes with only hardcore interactions}
\label{Appendix Positioning of multiple nucleosomes with only hardcore interactions}
Here we consider positioning of nucleosomes with hard-core interactions on up-exponential and down-exponential energy landscapes.

\subsubsection{Disordered energy landscape with \emph{down}-exponential distribution}
Consider positioning of $N$ nucleosomes on uncorrelated disordered energy profile down-exponentially distributed with some parameter $\mathcal{E}$ (see Eq.~\eqref{downexp}). In the regime $\mathcal{E} \ll 1$ the nucleosomes are poorly positioned, while in the opposite regime $\mathcal{E} \gg 1$, the positioning is good.

The derived requirement for positioning may sound weak. However, in fact it means that, say, for $\rho=70\%$ and $\mathcal{E}=1.5 k_\text{B}T$ (moderate positioning regime, $\mathcal{P} \simeq 0.6$) the typical energy well for a nucleosome is $7.5 \pm 2 k_\text{B}T$ deep (see Eq.~\eqref{E1downexp} with $L$ replaced by $L/N$), relative to a random DNA sequence.

\subsubsection{Disordered energy landscape with \emph{up}-exponential distribution}
Consider positioning of $N$ nucleosomes on uncorrelated disordered energy profile up-exponentially distributed with some parameter $\mathcal{E}$ (see Eq.~\eqref{upexp}). In this case, as for a single nucleosomes, the positioning is poor for any value of $\mathcal{E}$.

\subsection{Positioning of strongly interacting nucleosomes}
\label{Appendix Positioning of strongly interacting nucleosomes}
Here we discuss positioning of strongly interacting nucleosomes on up-exponential and down-exponential energy landscapes.

\subsubsection{Disordered energy landscape with \emph{down}-exponential distribution}
Consider positioning of $N$ nucleosomes on uncorrelated disordered energy profile down-exponentially distributed with some parameter $\mathcal{E}$ (see Eq.~\eqref{downexp}).
In this case a cluster of $m \gg 1$ crystallized nucleosomes has a Gaussian energy landscape with a standard deviation of $\sqrt{m}\mathcal{E}$. Using the same arguments as for the Gaussian disorder one can derive two conditions for a good positioning. The first is for positioning of weakly interacting nucleosomes, such that $M \simeq 1$. In this case the condition is given by Eq.~\eqref{ExpDownCond1}.

If this condition is not satisfied one needs strongly interacting nucleosomes, such that
\begin{equation}
	\sqrt{M} \mathcal{E} \gg \sqrt{2 \ln R}
\end{equation}
or, using Eq.~\eqref{M},
\begin{equation}
		\kappa \gg \frac{4 }{\mathcal{E}^4 }  \frac{L}{N}\ln^2 R.
		\label{ExpDownCond2}
\end{equation}
\begin{figure}[h!]
\centering
  \includegraphics[width=0.5\textwidth]{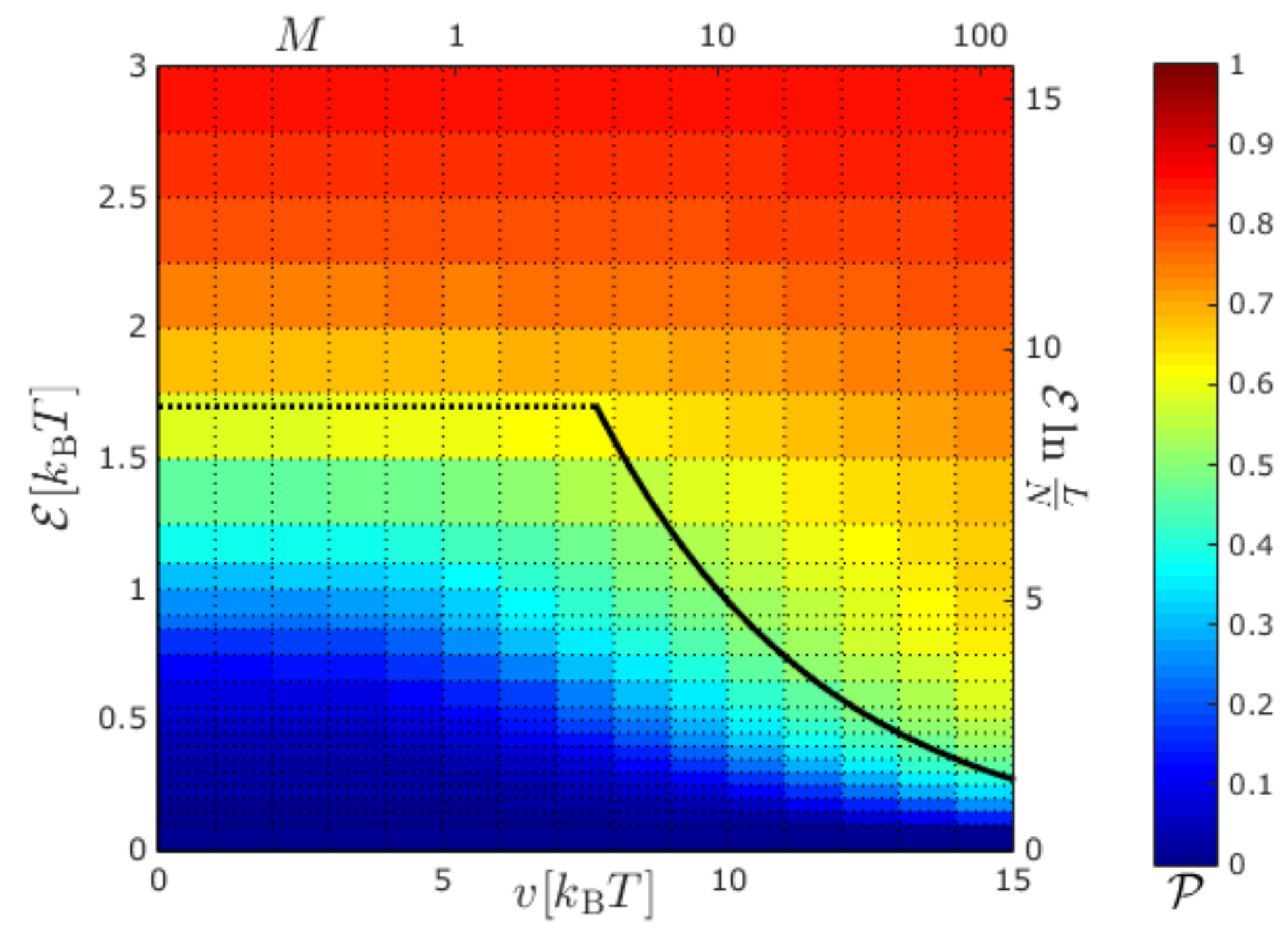}
\caption{Positioning of nucleosomes with interactions on down-exponential energy landscape. Positioning parameter, $\mathcal{P}$ for the DNA with length $L=10^4 \times W$ and $W=147$ for the coverage fraction $\rho=nW/L$ smaller than $80\%$ is plotted vs. disorder strength (left axis) and interaction strength (bottom axis) with preferable distance of $R=148$. On the top one can see the average size of the crystallized cluster of nucleosomes, derived from Eq.~\eqref{M}. On the right the typical binding energy of a nucleosome (relative to the average energy) is shown.
 The lines represent the analytic conditions for a good positioning, Eqs.~\eqref{ExpDownCond1} (dotted line) and \eqref{ExpDownCond2} (solid line).} 
\label{PhaseDiagramExpDown}
\end{figure}
\begin{figure}[h!]
\centering
  \includegraphics[width=0.5\textwidth]{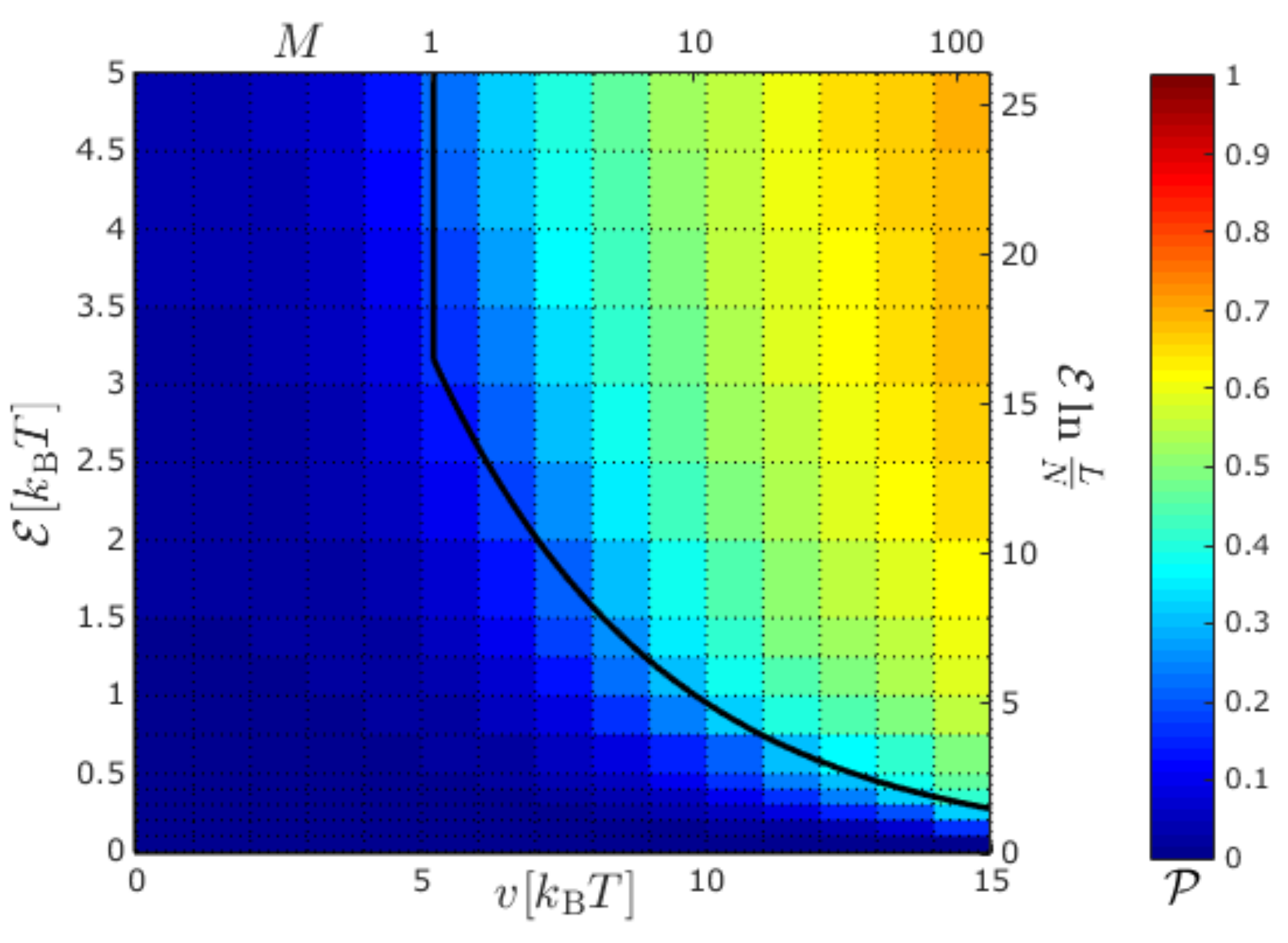}
\caption{Positioning of nucleosomes with interactions on up-exponential energy landscape. Positioning parameter, $\mathcal{P}$ for the DNA with length $L=10^4 \times W$ and $W=147$ for the coverage fraction $\rho=nW/L$ smaller than $80\%$ is plotted vs. disorder strength (left axis) and interaction strength (bottom axis) with preferable distance of $R=148$. On the top one can see the average size of the crystallized cluster of nucleosomes, derived from Eq.~\eqref{M}. On the right the typical energy \emph{peak} (not well) is shown on the length of $L/N$.
 The line represents the analytic conditions for a good positioning, Eq.~\eqref{ExpUpCond}.} 
\label{PhaseDiagramExpUp}
\end{figure}
\subsubsection{Disordered energy landscape with \emph{up}-exponential distribution}
Consider positioning of $N$ nucleosomes on uncorrelated disordered energy profile up-exponentially distributed with some parameter $\mathcal{E}$ (see Eq.~\eqref{upexp}). In this case, as for a single nucleosomes, the positioning is poor for any value of $\mathcal{E}$ and one needs strongly interacting nucleosomes for a good positioning.

If nucleosome strongly interact and form crystallized clusters of $M \gg 1$ nucleosomes on average (and this happens when $\kappa \gg \frac{L}{N}$), the positioning condition is given by Eq.~\eqref{ExpDownCond2}. Thus the condition for positioning in this case is
\begin{equation}
	\kappa \gg \frac{L}{N} \text{ and } \frac{4 }{\mathcal{E}^4 }  \frac{L}{N}\ln^2 R.
	\label{ExpUpCond}
\end{equation}

\bibliography{Ref}
\end{document}